\documentclass[12pt]{article}

\pdfoutput=1 
\usepackage{slashed}
\usepackage{color, verbatim}
\usepackage{latexsym}
\usepackage{amsmath}
\usepackage{graphicx}
\usepackage{float,lscape}
\usepackage{afterpage}

\usepackage{cite}
\usepackage{hyperref}

\usepackage{amssymb}
\usepackage{graphicx}
\usepackage{bm}
\makeatletter

\newcommand{\preprint}[1]{#1}

\@addtoreset{equation}{section}
\makeatother

\newcommand{\mc}[1]{#1}

\begin{document}
\preprint{IPPP/18/82}

\thispagestyle{empty}

\begin{center}

\begin{center}

\vspace{.5cm}

{\Large\sc Constraining four-fermion operators\\[0.5cm] using rare top decays}

\end{center}

\vspace{0.8cm}

\textbf{
Mikael Chala$^a$, Jose Santiago$^b$ and Michael Spannowsky$^a$}\\

\vspace{1.cm}
{\em {$^a$ Institute of Particle Physics Phenomenology, Physics Department, Durham University,
Durham DH1 3LE, UK}}\\[0.5cm]
{\em {$^b$ CAFPE and Departamento de F\'isica Te\'orica y del Cosmos, Universidad de Granada, E-18071 Granada, Spain}}

\end{center}

\begin{abstract}
New physics can manifest itself by an appreciable increase of the decay rate of top quarks in rare flavour-changing final states. Exploiting the large top quark production rate at the LHC, we bound four-fermion operators contributing to non-resonant $t\rightarrow \ell^+\ell^- j$ using different signal regions of the latest LHC searches for $t\rightarrow Zj$. We also provide prospects for the high-luminosity LHC to test these as well as four-fermion operators contributing to $t\rightarrow b\overline{b}j$, based on improved analysis strategies of existing searches. We single out all weakly-coupled ultraviolet completions inducing such contact interactions at tree level and translate the previous bounds to the parameter space of specific complete models. Being above the TeV, LHC bounds from rare top decays improve over those from flavour physics, electroweak precision data and other LHC searches in several cases.
\end{abstract}

\newpage

\tableofcontents

\newpage

\section{Introduction}
The very good agreement between the Standard Model (SM) predictions and the current data suggests that new physics might only lie at energies significantly above the electroweak (EW) scale. If this is true, its effects could be therefore accurately captured by a set of SM effective operators. One such independent operator exists at dimension five~\cite{Weinberg:1979sa}, whereas 59 independent operators up to flavour indices appear at dimension six~\cite{Buchmuller:1985jz,Grzadkowski:2010es}. Numerous studies have been performed with the aim of constraining the SM Effective Field Theory (SMEFT); see \textit{e.g.} Refs.~\cite{Buckley:2015lku, Englert:2015hrx, Butter:2016cvz, Falkowski:2017pss, Ellis:2018gqa, Fujii:2018mli} for recent analyses.

The operators with largest coefficients are expected to be those which can be induced at tree level in UV completions of the SM~\cite{deBlas:2017xtg}. Among them, we find in particular four-fermion operators. Studies of four-light-quark operators at the LHC can be found in Ref.~\cite{Domenech:2012ai}. Analyses of different types of four-lepton operators can be obtained in Refs.~\cite{delAguila:2014soa,delAguila:2015vza,Falkowski:2015krw,Falkowski:2017pss,Falkowski:2018dmy}. Likewise, bounds on two-light-quark-two-lepton operators using low-energy as well as LHC data have been obtained in Refs.~\cite{Cirigliano:2012ab,Carpentier:2010ue,deBlas:2013qqa,Falkowski:2015krw,Gonzalez-Alonso:2016etj,Farina:2016rws}. Searches for two-top-two-light-quark operators in top single and pair production have been worked out in Refs.~\cite{AguilarSaavedra:2010zi}.  Ref.~\cite{deBlas:2015aea} provided limits on four-top and two-top-two-lepton operators from their effects on EW Precision Data (EWPD) from Renormalization Group Evolution (RGE); see also Refs.~\cite{Bhattacharya:2018ryy,Falkowski:2017pss}. Studies of four-top operators have been also considered in four-top production~\cite{Degrande:2010kt} and top pair-production in association with $b$ quarks~\cite{Degrande:2010kt,Buckley:2015nca,DHondt:2018cww}. However, four-fermion operators involving one top as well as light quarks and/or leptons have received only little attention~\cite{Fox:2007in,Drobnak:2008br,AguilarSaavedra:2010zi,Durieux:2014xla,Kamenik:2018nxv}, despite appearing in several scenarios of new physics. In fact, no dedicated LHC search for these interactions has been developed. Analyses devoted for other experimental signatures might be sensitive to rare top decays mediated by four-fermion operators, though. Thus, in this article we consider the latest and most constraining search for $t\rightarrow Zj$ to date~\cite{Aaboud:2018nyl}, and demonstrate that it can already constrain the scale of contact interactions contributing to non-resonant $t\rightarrow \ell^+\ell^- j$ beyond the TeV. This possibility has been previously pointed out in the literature; see \textit{e.g.} Refs.~\cite{Fox:2007in,Durieux:2014xla}. However, actual limits on four-fermion interactions using existing searches for $t\rightarrow Zj$ have not been reported. Moreover, the reach of dedicated analyses for testing these operators at the LHC has not been estimated, which we intend to rectify. Thus, we also design new analyses to test four-fermion operators contributing to non-resonant $t\rightarrow b\overline{b} j$.

The reason for focusing on rare top decays is primarily motivated by the fact that top quarks are copiously produced at the LHC. Further, several contact-interactions, like those giving only $\sim t c \mu^+ \mu^-$ can not be directly probed otherwise.

This paper is organized at follows. In Section~\ref{sec:framework} we describe first the set of effective interactions we are interested in. We then focus on those contributing to $t\rightarrow \ell^+\ell^- j$. We recast the latest and most constraining search for $t\rightarrow Zj$ and provide bounds on the coefficients of such contact interactions. We also estimate the reach of modified versions of this analysis to constrain contact-interactions in the High-Luminosity LHC (HL-LHC) phase, defined by $\sqrt{s} = 14$ TeV and $\mathcal{L} = 3$ ab$^{-1}$. We subsequently focus on the $t\rightarrow b\overline{b} j$ channel, for which dedicated searches have not been performed yet.  We also comment briefly on the reach of other facilities to the contact-interactions we bound in this article in comparison with our findings. These include measurements of $b\to s\mu^+\mu^-$, $B_s$-$\overline{B_s}$ mixing, the $B_c^-$ lifetime, EWPD and the cross section of single-top production.

We extend the previous results to operators involving tau leptons as well as lepton-flavour violation (LFV) in Section~\ref{sec:beyond}. We compare the sensitivity of our analyses with the one achieved by low energy experiments such as $\mu^\pm\to e^\pm\gamma$, $\mu^\pm\to e^\pm e^\mp e^\pm$ and the tau counterparts. In Section~\ref{sec:matching} we provide the list of all possible weakly-coupled UV completions inducing the four-fermion operators of interest at tree level. We demonstrate that wide regions of their parameter spaces can be better bounded by searches for anomalous top decays than by other experiments. Finally, we conclude in Section~\ref{sec:conclusions}.

\section{Framework}
\label{sec:framework}
%
%\
%
The only four-fermion operators contributing to $t\rightarrow \ell^+\ell^- j$ are linear combinations of
\begin{equation}\label{eq:lepop1}
 \mathcal{O}_{lq}^{-(ijkl)} = \frac{1}{2}[\overline{l^i_L}\gamma^\mu l^j_L)(\overline{q^k_L}\gamma_\mu q^l_L)-(\overline{l^i_L}\gamma^\mu\sigma^I l^j_L)(\overline{q^k_L}\gamma_\mu\sigma_I q^l_L)]~,  \mathcal{O}_{eq}^{(ijkl)} = (\overline{e^i_R}\gamma^\mu e^j_R)(\overline{q^k_L}\gamma_\mu q^l_L)~,
 \end{equation}
 \begin{equation}\label{eq:lepop2}
 \mathcal{O}_{lu}^{(ijkl)} = (\overline{l^i_L}\gamma^\mu l^j_L)(\overline{u^k_R}\gamma_\mu u^l_R)~, \quad  \mathcal{O}_{eu}^{(ijkl)} = (\overline{e^i_L}\gamma^\mu e^j_L)(\overline{u^k_R}\gamma_\mu u^l_R)~,
 \end{equation}
 \begin{equation}\label{eq:lepop3}
 \mathcal{O}_{lequ}^{1(ijkl)} = (\overline{l_L^i} e_R^j) ~\varepsilon~ (\overline{q_L^k}u_R^l)~, \quad \mathcal{O}_{lequ}^{3(ijkl)} = (\overline{l_L^i} \sigma_{\mu\nu} e_R^j) ~\varepsilon~ (\overline{q_L^k} \sigma^{\mu\nu} u_R^l)~,
\end{equation}
 where $i,j,k,l$ are flavour indices (one of the quark flavour indices
 will correspond to the third generation and the other to one of the
 first two), $\sigma^I$ are the Pauli
 matrices and $\varepsilon \equiv \mathrm{i}\sigma^2$.
%% Assuming no LFV, namely $i=j$, we can define
%% $\mathcal{O}_{lq}^{-(a+3)} \equiv \mathcal{O}_{lq}^{-(iia3)} +
%% \mathcal{O}_{lq}^{-(ii3a)}$ with $a=1$ $(2)$ for up (charm) quarks; as
%% well as the operators $\mathcal{O}_{LQ}^{(a+3)} \equiv
%% \mathcal{O}_{LQ}^{iia3} + \mathcal{O}_{LQ}^{ii3a}$, with $LQ = eq, lu,
%% eu$. Note that all these operators are hermitian. Likewise, we
%% consider the non-hermitian operators $\mathcal{O}_{lequ}^{1(a3)}
%% \equiv \mathcal{O}_{lequ}^{1(iia3)}$ and $\mathcal{O}_{lequ}^{1(3a)}
%% \equiv \mathcal{O}_{lequ}^{1(ii3a)}$ as well as
%% $\mathcal{O}_{lequ}^{3(a3)} \equiv \mathcal{O}_{lequ}^{3(iia3)}$ and
%% $\mathcal{O}_{lequ}^{3(3a)} \equiv
%% \mathcal{O}_{lequ}^{3(ii3a)}$. These are the only operators
%% contributing to non-resonant $t\rightarrow \ell^+ \ell^- j$ at
%% dimension six.   

%

Analogously, in the four-quark sector we consider the following set of
linearly-independent operators:  
\begin{align}
 \mathcal{O}_{qq}^{1(ijkl)} & = (\overline{q_L^i}\gamma^\mu
 q_L^j)(\overline{q_L^k}\gamma_\mu q_L^l)~,
 &\mathcal{O}_{qq}^{3(ijkl)} &= (\overline{q_L^i}\sigma^I\gamma^\mu
 q_L^j)(\overline{q_L^k}\sigma_I\gamma_\mu q_L^l)~, \label{eq:hadop1}\\ 
 \mathcal{O}_{qu}^{1(ijkl)} &= (\overline{q_L^i}\gamma^\mu
 q_L^j)(\overline{u_R^k}\gamma_\mu u_R^l)~,
 &\mathcal{O}_{qu}^{8(ijkl)} &= (\overline{q_L^i}T^A\gamma^\mu
 q_L^j)(\overline{u_R^k}T_A\gamma_\mu u_R^l)~,\\ 
 \mathcal{O}_{qd}^{1(ijkl)} &= (\overline{q_L^i}\gamma^\mu
 q_L^j)(\overline{d_R^k}\gamma_\mu d_R^l)~,
 &\mathcal{O}_{qd}^{8(ijkl)} &= (\overline{q_L^i}T^A\gamma^\mu
 q_L^j)(\overline{d_R^k}T_A\gamma_\mu d_R^l)~,\\ 
 \mathcal{O}_{ud}^{1(ijkl)} &= (\overline{u_R^i}\gamma^\mu
 u_R^j)(\overline{d_R^k}\gamma_\mu d_R^l)~,
 &\mathcal{O}_{ud}^{8(ijkl)} &= (\overline{u_R^i}T^A\gamma^\mu
 u_R^j)(\overline{d_R^k}T_A\gamma_\mu d_R^l)~,\\ 
 \mathcal{O}_{quqd}^{1(ijkl)} &= (\overline{q_L^i} u_R^j)
 ~\varepsilon~ (\overline{q_L^k}^T d_R^l)~,
 &\mathcal{O}_{quqd}^{8(ijkl)} &= (\overline{q_L^i} T^A u_R^j)
 ~\varepsilon~ (\overline{q_L^k}^T T_A d_R^l)~. \label{eq:hadop2}
\end{align}
where $T_A = \lambda^A/2$ with $\lambda^A$ the Gell-Mann matrices.
%% Out of those operators we construct the hermitian operators
%% $\mathcal{O}_{QQ}^{I(a+3)}\equiv \mathcal{O}_{QQ}^{I(333a)} +
%% \mathcal{O}_{QQ}^{I(33a3)}$, with $QQ (I) = qq (1)$, $qq (3)$, $qu
%% (1)$, $qu (8)$ and $\mathcal{O}_{QQ}^{I(a+3)}\equiv
%% \mathcal{O}_{QQ}^{I(3a33)} + \mathcal{O}_{QQ}^{I(a333)}$ with $QQ (I)
%% = qd (1), qd (8), ud (1), ud (8)$. 
%% %
%% %
%% Besides, we consider the non-hermitian operators $\mathcal{O}_{quqd}^{1(ijkl)} \equiv (\overline{q_L^i} u_R^j)\varepsilon (\overline{q_L^k}d_R^l)$ and $\mathcal{O}_{quqd}^{8(ijkl)} \equiv (\overline{q_L^i}T^A u_R^j)\varepsilon (\overline{q_L^k} T_A d_R^l)$.
%% These are the only dimension-six operators contributing to non-resonant $t\rightarrow b\overline{b} j$.

When giving specific numerical results in this section we will
consider the case $j = c, \ell^\pm = \mu^\pm$. 
Implications of departures from this assumption will be
dicussed in Section~\ref{sec:beyond}. 

%
%.

\subsection{Results for $t\rightarrow\ell^+\ell^- j$}
The relevant effective Lagrangian is given by
\begin{align}
  \mathcal{L}= \frac{1}{\Lambda^2}
&  \bigg[
    c^{-\,(ijkl)}_{lq} \mathcal{O}^{-\,(ijkl)}_{lq}
   +c^{(ijkl)}_{eq} \mathcal{O}^{(ijkl)}_{eq}
   +c^{(ijkl)}_{lu} \mathcal{O}^{(ijkl)}_{lu}
   +c^{(ijkl)}_{eu} \mathcal{O}^{(ijkl)}_{eu}
   \nonumber \\
   &+\Big\{
    c^{1(ijkl)}_{lequ} \mathcal{O}^{1(ijkl)}_{lequ}
   +c^{3(ijkl)}_{lequ} \mathcal{O}^{3(ijkl)}_{lequ}
   +\mathrm{h.c.}\Big\}
    \bigg].
  \end{align}
  The decay width of the top quark from this effective Lagrangian was
  computed in Ref.~\cite{AguilarSaavedra:2010zi} in a diferent
  basis. Translating it to our operator basis we get
  %\mc{[I shited the second row to the left for the equation to number to be in place]}
\begin{align}
  \Gamma(t\to \ell^+_i \ell^-_j u_k )= &\frac{m_t}{6144\pi^3} \left(
  \frac{m_t}{\Lambda}\right)^4 \Big\{ 4 |c^{-(jik3)}_{lq}|^2 + 4
  |c^{(jik3)}_{eq}|^2
  +4 |c^{(jik3)}_{lu}|^2 + 4 |c^{(jik3)}_{eu}|^2 \nonumber \\
  +& 
  |c^{1\,(jik3)}_{lequ}|^2+|c^{1\,(ij3k)}_{lequ}|^2
  +48  |c^{3\,(jik3)}_{lequ}|^2+ 48 |c^{3\,(ij3k)}_{lequ}|^2\Big\}.\label{eq:lepwidth}
  \end{align}
Since we are only sensitive to the absolute value of the Wilson
coefficients, we assume in the following that they are real.
In particular this implies that
\begin{align}
  c^{-(jilk)}_{lq}=c^{-(ijkl)}_{lq},\quad
  c^{(jilk)}_{eq}=c^{(ijkl)}_{eq},\quad
  c^{(jilk)}_{lu}=c^{(ijkl)}_{lu},\quad
  c^{(jilk)}_{eu}=c^{(ijkl)}_{eu},
\end{align}
%% We consider the Lagrangian
%% %
%% \begin{align}\label{eq:lepop}\nonumber
%%  L = \frac{1}{\Lambda^2}\bigg[&c_{l q}^{-(a+3)} \mathcal{O}_{l q}^{-(a+3)} + c_{eq}^{(a+3)} \mathcal{O}_{eq}^{(a+3)} + c_{l u}^{(a+3)} \mathcal{O}_{l u}^{(a+3)} + c_{eu}^{(a+3)} \mathcal{O}_{eu}^{(a+3)}   \\
%%  &+ \bigg\lbrace c_{lequ}^{1 (a3)} \mathcal{O}_{lequ}^{1 (a3)} + c_{lequ}^{1 (3a)} \mathcal{O}_{lequ}^{1 (3a)} + c_{lequ}^{3 (a3)} \mathcal{O}_{lequ}^{3 (a3)} + c_{lequ}^{3 (3a)} \mathcal{O}_{lequ}^{3 (3a)} + \text{h.c.}\bigg\rbrace\bigg]~,
%% \end{align}
%% %
%% for which we use the shorten notation introduced above.
%% %
%% We assume all coefficients to be real. The Lagrangian above is hermitian by construction. %T
%% Using \texttt{MadGraph v5}~\cite{Alwall:2014hca} with the \texttt{UFO} model~\cite{Degrande:2011ua} of Ref.~\cite{AguilarSaavedra:2018nen} at leading order, we obtain:
%% %
%% \begin{align}\label{eq:lepwidth}
%%  \Gamma(t\rightarrow \ell^+ \ell^- j) &= \frac{10^{-6}}{\Lambda^4}\bigg[ 3.1\times\bigg\lbrace |c_{l q}^{-(a+3)}|^2 + |c_{eq}^{(a+3)}|^2 + |c_{l u}^{(a+3)}|^2 + |c_{eu}^{(a+3)}|^2 \bigg\rbrace\\\nonumber
%%  &+  0.80\times \bigg\lbrace |c_{lequ}^{1 (a3)}|^2 + |c_{lequ}^{1 (3a)}|^2 \bigg\rbrace + 37\times\bigg\lbrace |c_{lequ}^{3 (a3)}|^2 + |c_{lequ}^{3 (3a)}|^2 \bigg\rbrace\bigg]~\text{GeV}~\text{TeV}^4~.
%% \end{align}
%
We note that 
\begin{align}\label{eq:lepxsec}\nonumber
 \sigma(p p \rightarrow t\overline{t}, t (\overline{t})\rightarrow \ell^+\ell^- j, \overline{t}(t)\rightarrow\text{all}) &= 2\times \sigma(pp\rightarrow t\overline{t})\times\mathcal{B}(t\rightarrow\ell^+\ell^- j)~\\
 &=2\times \sigma(pp\rightarrow t\overline{t})\times\frac{\Gamma(t\rightarrow \ell^+ \ell^- j)}{\Gamma_t}~,
\end{align}
where the top quark's total width is $\Gamma_t\sim 1.35$ GeV and
$\sigma(pp\rightarrow t\overline{t}) \sim 830$ pb at
NNLO~\cite{kfactors1}. 
Using the production cross section in terms of the partial width
and the fact that the number of
events in a certain signal region of an analysis is given by
\begin{equation}
  s=\sigma \times \epsilon \times
  \mathcal{L}, \label{eq:signal:general}
\end{equation}
where $\epsilon$ is the efficiency for the signal in the corresponding
region and $\mathcal{L}$ the integrated luminosity used in the analysis, we can
write a master equation for the observed number of signal events in
specific regions of parameter space:
\begin{align}
  s(t \to \ell^+_i \ell^-_j u_k)=
  \frac{1}{\Lambda^4}
  \Big[ &
    \alpha_{lq}^{-(jik3)} |c^{-(jik3)}_{lq}|^2 +
    \alpha_{eq}^{(jik3)} |c^{(jik3)}_{eq}|^2 \nonumber \\
    +&
\alpha_{lu}^{(jik3)} |c^{(jik3)}_{lu}|^2 +
\alpha_{eu}^{(jik3)} |c^{(jik3)}_{eu}|^2 \nonumber \\
+&
\alpha_{lequ}^{1(jik3)} |c^{1(jik3)}_{lequ}|^2 +
\alpha_{lequ}^{1(ij3k)} |c^{1(ij3k)}_{lequ}|^2 \nonumber \\
+&
\alpha_{lequ}^{3(jik3)} |c^{3(jik3)}_{lequ}|^2 +
\alpha_{lequ}^{3(ij3k)} |c^{3(ij3k)}_{lequ}|^2\Big], \label{eq:masterllj}
\end{align}
where the different $\alpha$ encode the efficiencies of the particular
analysis for each contribution.

The LHC has the largest sensitivity to several of the four-fermions operators above~\cite{Fox:2007in}. However, no dedicated analysis in this respect has been worked out yet. Among the most constraining analyses we find therefore searches for $t\rightarrow Z j$, mediated by operators such as $\mathcal{O}_{uB}^{(ij)} = (\overline{q^i_L}\sigma^{\mu\nu} u_R^j)\tilde{\varphi}B_{\mu\nu}$, with $\varphi$ the Higgs doublet. Although the leptons resulting from the top decay via the contact interactions do not always reconstruct a $Z$ boson, we will show that these kind of searches do still have a large sensitivity. %
To this end, we consider the latest ATLAS search~\cite{Aaboud:2018nyl} for FCNC top quark decays at LHC13, \textit{i.e.} $\sqrt{s} = 13$ TeV and $\mathcal{L} = 36$ fb$^{-1}$. It provides the strongest limit on $\mathcal{B}(t\rightarrow Zj)$ to date. In short terms, this analysis demands three light leptons (either electrons or muons), two of them same-flavour opposite-sign (SFOS), as well as exactly one $b$-tagged jet and at least two more light jets. (The $b$-tagging efficiency is reported to be $0.77$, while the misstag rates $\mathcal{P}(c\rightarrow b)$ and $\mathcal{P}(j\rightarrow b)$ are $0.16$ and $0.0075$ respectively~\cite{Aaboud:2018nyl}.)

\begin{figure}[t]
 \begin{center}
  \includegraphics[width=0.487\columnwidth]{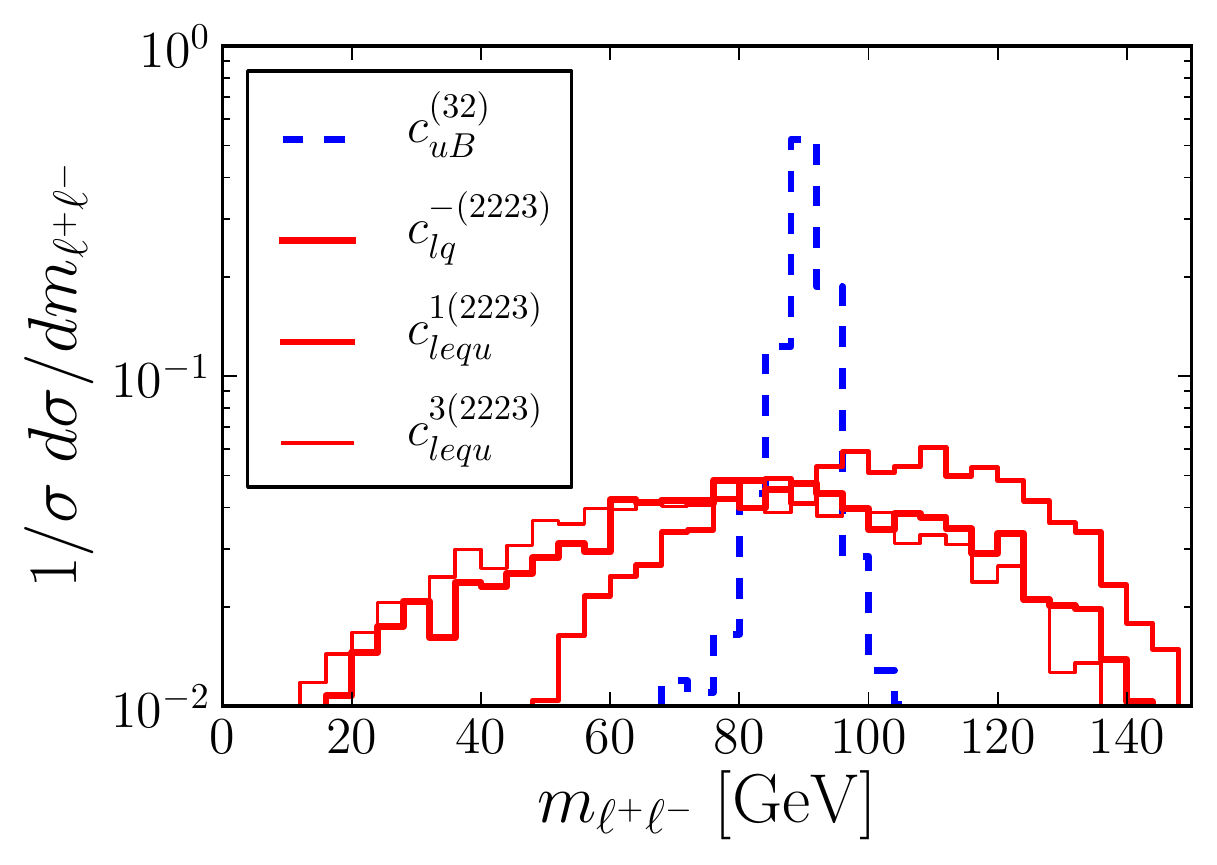}
  \includegraphics[width=0.503\columnwidth]{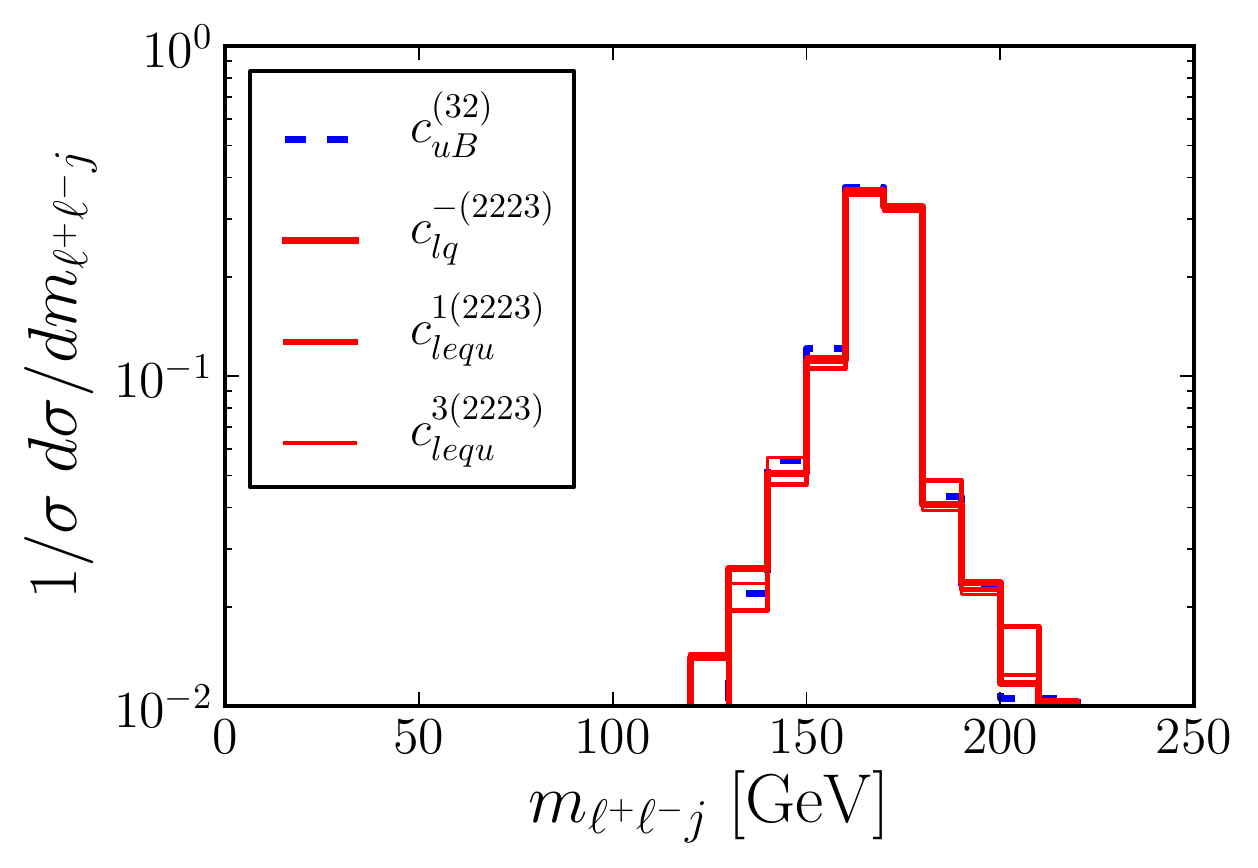}
 \end{center}
\caption{Distribution of $m_{\ell^+\ell^-}$ (left) and $m_{\ell^+\ell^- j}$ (right) after the basic cuts of Ref.~\cite{Aaboud:2018nyl} for four different operators.}\label{fig:histo}
\end{figure}

The two SFOS leptons whose invariant mass $m_{\ell^+\ell^-}$ is closest to the 
$Z$ pole are considered as the $Z$ boson candidate.
\mc{We notice that most events peak at invariant masses very different from the 
$Z$ mass in the 
non-resonantly produced signal.} However, a sizeable fraction of the events 
still populate the \mc{$Z$ peak} despite the leptons being produced from 
effective operators rather that from the decay of the $Z$. This result is 
explicitly shown in the left panel of Fig.~\ref{fig:histo}, where the 
distribution of $m_{\ell^+\ell^-}$ after the basic cuts of the experimental 
analysis is depicted. Moreover, the total transverse missing energy is forced to 
be $E_T^\text{miss} > 20$ GeV.

On top of the observables above, the analysis considers the invariant masses $m_{\ell^+\ell^- j}$, $m_{\ell^\pm b \nu}$ and $m_{\ell^\pm \nu}$, where $\ell^\pm$ refers to the non-$Z$ lepton, $b$ is the $b$-tagged jet and $\nu$ the neutrino. The momentum of the latter, as well as the momentum of the selected jet $j$ are those minimizing
\begin{equation}
 \chi^2 = \frac{(m_{\ell^+\ell^- j} - m_{t_{\text{FCNC}}})^2}{\sigma_{t_{\text{FCNC}}}^2} + \frac{(m_{\ell^\pm b \nu} - m_{t_{\text{SM}}})^2}{\sigma_{t_{\text{SM}}}} + \frac{(m_{\ell^\pm \nu} - m_W)^2}{\sigma_W}~,
\end{equation}
with $m_{t_{\text{FCNC}}} = 169.6$ GeV, $m_{t_{\text{SM}}} = 167.2$ GeV, $m_W = 81.2$ GeV, $\sigma_{t_{\text{FCNC}}} = 12.0$ GeV, $\sigma_{t_{\text{SM}}} = 24.0$ GeV and $\sigma_W = 15.1$ GeV. These variables behave almost equally in the $t\rightarrow Zj$ and the contact-interaction cases; see an example in the right panel of Fig.~\ref{fig:histo}.

Two regions of interest for our analysis are studied in the experimental paper. 
% \mc{[FOLLOWING THE REFEREE'S SUGGESTION, WE SHOULD DISREGARD THE SRA AS WELL AS 
% TAKING THE EVENTS IN SRB FROM A DIFFERENT TABLE THAT GIVES RAW DATA INSTEAD OF 
% FIT ONE.]}
First, the signal region SRA. It requires $m_{\ell^+\ell^-}$ to be 
within the range $[76.2, 106.2]$ GeV. It further imposes $|m_{\ell^+\ell^- j} - 
172.5| < 40$ GeV, $|m_{\ell^\pm b \nu} - 172.5| < 40$ GeV and $|m_{\ell^\pm \nu} 
- 80.4| < 30$ GeV. \mc{We use this region to validate our simulations}. Given 
the numbers in Table~8 of the experimental reference, one can conclude that the 
efficiency for selecting $t\overline{t}$ events with one of the top decaying as 
$t\rightarrow Zc$ is about $2.4$~\%.

We have recast the corresponding cuts using home-made routines based
on \texttt{ROOT v6}~\cite{Brun:1997pa} and \texttt{Fastjet
  v3}~\cite{Cacciari:2011ma}. We have applied them to Monte Carlo
events generated using \texttt{MadGraph v5}~\cite{Alwall:2014hca} with
the \texttt{UFO} model~\cite{Degrande:2011ua} of
Ref.~\cite{AguilarSaavedra:2018nen}. We have further used
\texttt{Pythia v6}~\cite{Sjostrand:2006za} to simulate radiation,
fragmentation and hadronization processes. The efficiency we obtain
matches very well that previously reported. 
\begin{table}[t]
\footnotesize
\begin{center}
 \begin{tabular}{|l|cccccccc|}
 \hline
 						  &			 &			&		   &		      &					  & & &\\[-0.1cm]
  & $\alpha_{l q}^{-(2223)}$ & $\alpha_{e q}^{(2223)}$ & $\alpha_{l u}^{(2223)}$ & $\alpha_{e u}^{(2223)}$ & $\alpha_{l equ}^{1(2223)}$ & $\alpha_{l equ}^{1(2232)}$ & $\alpha_{l equ}^{3(2223)}$ & $\alpha_{l equ}^{3(2232)}$ \\[0.2cm]
  \hline
  						  &			 &			&		   &		      &					  & & &\\[-0.1cm]
  CR1  & 2.0   & 2.0 & 2.0  & 2.0  & 0.44  & 0.44 & 26.0  & 26.0 \\[0.2cm]
  NEW  & 1.8   & 1.8 & 1.8  & 1.8  & 0.37  & 0.37 & 23.0  & 23.0 \\[0.2cm]
  \hline
 \end{tabular}
\caption{ Coefficients of the master equation (\ref{eq:masterllj}), in $\mathrm{TeV}^4$,
  for the different signal regions and for LHC13. In the HL-LHC,
  the coefficients should be multiplied by a factor 100 to account for the
  increase in production cross section and luminosity. See text for
  the definition of the different signal regions.} 
\label{tab:masterlljcoeffs}
\end{center}
\end{table}

Another region of interest, not used in the experimental analysis to
bound new physics but rather to validate the non-prompt lepton
backgrounds, is the one dubbed CR1~\cite{Aaboud:2018nyl}. On top of
the basic cuts, it also requires $|m_{\ell^+\ell^-}-91.2| > 15$ GeV,
being therefore more sensitive to four-fermion operators. It does not
cut on $m_{\ell^+\ell^- j}$, $m_{\ell^\pm b\nu}$ or
$m_{\ell^\pm\nu}$. We have computed the corresponding efficiencies,
obtaining \mc{values} $\sim 0.015$.
% for
% $\mathcal{O}_{lq}^{-}$, $\mathcal{O}_{eq}$,
% $\mathcal{O}_{lu}$ and $\mathcal{O}_{eu}$; $0.013$
% for $\mathcal{O}_{lequ}^{1}$; and $0.016$ for
% $\mathcal{O}_{lequ}^{2}$.
This results in the coefficients of the
master equation shown in the \mc{first} row of Table~\ref{tab:masterlljcoeffs}.
The experimental collaboration reports the observation of
260 observed events, while \mc{$230\pm 70$} are predicted. 
\mc{Using the CL$_s$ method~\cite{Read:2002hq}, we determine the maximum number 
of signal events allowed at the 95 \% CL including the $\sim 30\,\%$ 
systematic uncertainty on the SM prediction, obtaining
$s_{max} = 143$}. 
% .
\mc{Our master equation then allows us to set limits on arbitrary
combinations of the coefficients of the four-fermion operators.
Assuming for simplicity one operator at a time these bounds are shown,
for $\Lambda = 1$ TeV, in boldface in the first row of
Table~\ref{tab:lepbounds}.
}

\mc{We also provide naive prospects for the
HL-LHC. For this aim, we scale the background cross section by a
conservative factor of $1.3$. This number corresponds to the
enhancement in cross section for $t\overline{t}Z$ from $\sqrt{s} =
13$ TeV to $\sqrt{s} = 14$ TeV, being the largest among the
dominant backgrounds. For the signal, we assume an enhancement of
$\sim 1.2$~\cite{kfactors1}. We further scale the number of events
with the ratio of luminosities, $\sim
3~\text{ab}^{-1}/36~\text{fb}^{-1}\sim 83$. 
%M.
Assuming the number of observed events equal to the number of SM
events, we find that $s_{max} = 315$. \mc{(This number 
becomes an order of magnitude larger for systematic uncertainties 
of $10$ \%}.) 
% 
%	\ms{XXX}. %, depending on the operator.
Projected bounds on the
operator coefficients are also shown, within parentheses, in the first
row of Table~\ref{tab:lepbounds}. 
}

We can improve further these bounds by extending CR1 with the cuts on $m_{\ell^+\ell^- j}$, $m_{\ell^\pm b \nu}$ and $m_{\ell^\pm \nu}$ required in SRA. Such a new sharpened signal region has not been yet considered experimentally.
\begin{table}[t]
\footnotesize
\begin{center}
%  
%  \end{tabular}
 \begin{tabular}{|l|cccccccc|}
 \hline
 						  &			 &		
	&		   &		      &					 
 & & &\\[-0.1cm]
  & $c_{l q}^{-(2223)}$ & $c_{e q}^{(2223)}$ & $c_{l u}^{(2223)}$ & $c_{e 
u}^{(2223)}$ & $c_{l equ}^{1(2223)}$ & $c_{l equ}^{1(2232)}$ & $c_{l 
equ}^{3(2223)}$ & $c_{l equ}^{3(2232)}$ \\[0.2cm]
  \hline
  						  &			 &		
	&		   &		      &					 
 & & &\\[-0.1cm]
  CR1  & $\mathbf{8.4}$ $(1.2)$ & $\mathbf{8.4}$ $(1.2)$ & $\mathbf{8.4}$ 
$(1.2)$ & $\mathbf{8.4}$ $(1.2)$ & $\mathbf{18}$ $(2.7)$ & $\mathbf{18}$ 
$(2.7)$ & $\mathbf{2.3}$ $(0.35)$ & $\mathbf{2.3}$ $(0.35)$\\[0.2cm]
  NEW  & $3.1$ $(1.0)$ & $3.1$ $(1.0)$ & $3.1$ $(1.0)$ & $3.1$ $(1.0)$ & $6.8$ 
$(2.2)$ & $6.8$ $(2.2)$ & $0.87$ $(0.28)$ & $0.87$ $(0.28)$\\[0.2cm]
  \hline
 \end{tabular}
\caption{Bounds on $c$ for $\Lambda = 1$ TeV, asuming one operator at
  a time, using the different signal regions defined in the text. 
The numbers without (within) parenthesis stand for the LHC13
(HL-LHC).
The boldface indicates limits using actual data. These numbers can be
obtained from the master equation (\ref{eq:masterllj}) using the
coefficients in Table~\ref{tab:masterlljcoeffs} and the upper bound on
the following number of signal events:
%$s^{SRA}_{\mathrm{max}}=21~(225)$,
$s^{CR1}_{\mathrm{max}}=143~(315)$ and
$s^{NEW}_{\mathrm{max}}=18~(179)$, where again the number in brackets
correspond to HL-LHC projections. \mc{The projected bounds on the 
coefficients get a factor of $\sim 3$ weaker for systematic uncertainties of 
$10\,\%$.}
}\label{tab:lepbounds}
\end{center}
\end{table}
Therefore, we estimate the number of expected SM events from simulation. To this end, we first check our Monte Carlo for the background comparing the expectations for CR1 with those reported in the experimental analysis.
This region is dominated, first, by non-prompt leptons (coming mainly
from $t\overline{t}$). We get an efficiency of selecting events in the
CR1 region of $\sim 0.04$. Fixing the misstag rate
$\mathcal{P}(j\rightarrow e^\pm) = 3\times 10^{-4}$ (which is well
within the actual range; see \textit{e.g.}
Ref.~\cite{daSilvaFernandesdeCastro:2008zz}), we match the number of
events reported in \mc{Table 5 of Ref.~\cite{Aaboud:2018nyl}: $140$ ($\pm
70$)}. We find good agreement in the other backgrounds too, with the
exception of $WZ$ for which we have much less events than provided in
the paper. Still, being subdominant, and given the large error
reported by ATLAS in the determination of that background, we proceed
without this sample. Around $\sim 73$ background events survive in
this new region for $\mathcal{L} = 36$ fb$^{-1}$. Assuming the
observation of only background events \mc{and no systematic uncertainties}, we 
get $s_{max} \sim 18$ ($\sim
179$ for the HL-LHC). \mc{These numbers get a factor of $\sim 1.2$ ($\sim 10$)
larger with $10 \%$ systematics.} The signal efficiencies
are in this case of about $\sim 0.013$.
The corresponding coefficients for the
master equation are given
in the \mc{second} row of Table~\ref{tab:masterlljcoeffs}.
Using these numbers, we
obtain the limits as before and show them in the last row on
Table~\ref{tab:lepbounds}. Being non-boldface, we emphasize again that
they are not actual limits but potential ones. 

We note that scales as large as $\sim 1$ TeV for $\mathcal{O}(1)$
couplings are already bounded in some cases. The best limits at the
HL-LHC will probe scales of the order of $\Lambda\sim 2$ $(3.5)$ TeV
for $c\sim 1$ ($\sqrt{4\pi}$).
%. 
\mc{Bounds from flavour physics are more stringent}  for operators
involving Left-Handed (LH) quarks, whereas they are irrelevant for
Right-Handed (RH) ones. Indeed, in the former case $b\to s\mu^+\mu^-$
transitions arise at tree level. They modify the $\mathcal{B}(B_s\to
\mu^+\mu^-)$ by an amount of $\sim g_2^4/(16 \pi \Lambda^4) f_B^2
m_\mu^2 m_B/\Gamma_B$, with $g_2$ the $SU(2)_L$ gauge coupling,
$f_B\sim 0.2$ GeV, $m_B\sim 5$ GeV and $\Gamma_B\sim 4\times
10^{-13}$~\cite{Cheung:2006tm}. Therefore, for $\Lambda = 1$ TeV,
$\mathcal{B}(B_s\to \mu^+\mu^-)\sim 6\times 10^{-6}$, orders of
magnitude larger than the measured value $(2.8^{+ 0.7}_{-0.6})\times
10^{-9}$~\cite{CMS:2014xfa}. 

However, the contribution of operators such as
$\mathcal{O}_{eu}^{(2223)}$ is chirality and loop suppressed, being
therefore further reduced by a factor of $\sim N_c^2 g_2^4/(16
\pi^2)^2 m_c^2/v^2\sim 10^{-10}$ and hence negligible. The decay of
the vector meson $B_s^*\rightarrow\mu^+\mu^-$ could be much larger
because it is not helicity
suppressed~\cite{Grinstein:2015aua,Sahoo:2016edx}. However, to the
date there are no direct measurements of this observable.

\subsection{Results for $t\rightarrow b\overline{b} j$}
The relevant Lagrangian reads for this case
\begin{align}\label{eq:qukop}\nonumber
\mathcal{L} = \frac{1}{\Lambda^2}&\bigg[c_{qq}^{1(ijkl)}
  \mathcal{O}_{qq}^{1(ijkl)} +
  c_{qq}^{3(ijkl)}\mathcal{O}_{qq}^{3(ijkl)} +
  c_{qu}^{1(ijkl)}\mathcal{O}_{qu}^{1(ijkl)} +
  c_{qu}^{8(ijkl)}\mathcal{O}_{qu}^{8(ijkl)} \nonumber \\
  +&
  c_{qd}^{1(ijkl)}\mathcal{O}_{qd}^{1(ijkl)}
  + c_{qd}^{8(ijkl)}\mathcal{O}_{qd}^{8(ijkl)}
  +c_{ud}^{1(ijkl)}\mathcal{O}_{ud}^{1(ijkl)}+
  c_{ud}^{8(ijkl)}\mathcal{O}_{ud}^{8(ijkl)} 
  \nonumber \\+&  \bigg\lbrace
  c_{quqd}^{1(ijkl)}\mathcal{O}_{ququ}^{1(ijkl)} +
  c_{quqd}^{8(ijkl)}\mathcal{O}_{ququ}^{8(ijkl)}
 %% ~\\
%%  %& ~\\
%% & +
%%   
  + \text{h.c.}\bigg\rbrace\bigg]~.
\end{align}
%
%
%
%
%T
These operators alter the width of $\Gamma(t\to b\overline{b}j)$ at
the level of $1/\Lambda^4$. The only exceptions are the LL operators,
which interfere with the SM. However, the interference is suppressed
by a factor of $\sim V^{\text{CKM}}_{12}/100$; see Eq.~(21) in
Ref.~\cite{AguilarSaavedra:2010zi}, and we will neglect it in the rest
of this article. Translating the results
of~\cite{AguilarSaavedra:2010zi} to our basis we obtain the following
expression for the decay width:
%, 
\begin{align}
\Gamma(t\to b \bar{b} u_i)= \frac{m_t}{2048
  \pi^3}\left(\frac{m_t}{\Lambda}\right)^4
& \bigg\{
4 \Big[
  |c_{qq}^{1(33i3)}|^2
+ |c_{qu}^{1(33i3)}|^2
+ |c_{qd}^{1(i333)}|^2
+ |c_{ud}^{1(i333)}|^2
\Big]
\nonumber \\&
+
\frac{8}{9} \Big[
  \frac{33}{2}|c_{qq}^{3(33i3)}|^2
+ |c_{qu}^{8(33i3)}|^2
+ |c_{qd}^{8(i333)}|^2
+ |c_{ud}^{8(i333)}|^2
\Big]
\nonumber \\&
-\frac{8}{3} \mathrm{Re}[(c_{qq}^{1(33i3)})(c_{qq}^{3(33i3)})^\ast]
\nonumber \\&
+|c_{quqd}^{1(i333)}|^2
+|c_{quqd}^{1(33i3)}|^2
+\frac{7}{3}|c_{quqd}^{1(3i33)}|^2
\nonumber \\&
+\frac{2}{9}\big(|c_{quqd}^{8(i333)}|^2
+|c_{quqd}^{8(33i3)}|^2\big)
+\frac{10}{27}|c_{quqd}^{8(3i33)}|^2
\nonumber \\&
+\frac{1}{3}\mathrm{Re}[(c_{quqd}^{1(i333)})(c_{quqd}^{1(33i3)})^\ast]
-\frac{2}{27}\mathrm{Re}[(c_{quqd}^{8(i333)})(c_{quqd}^{8(33i3)})^\ast]
\nonumber \\&
+\frac{4}{9}\mathrm{Re}[(c_{quqd}^{1(i333)})(c_{quqd}^{8(33i3)})^\ast]
+\frac{4}{9}\mathrm{Re}[(c_{quqd}^{8(i333)})(c_{quqd}^{1(33i3)})^\ast]
\nonumber \\&
+\frac{8}{9}\mathrm{Re}[(c_{quqd}^{1(3i33)})(c_{quqd}^{8(3i33)})^\ast]
\bigg\}.
  \label{eq:hadwidth}
\end{align}
Using this expression and the equivalent of Eq.~(\ref{eq:lepxsec}) for
the hadronic case and Eq.~(\ref{eq:signal:general}) we can write a
master equation for the number of signal events
\begin{align}
  s(t\to b \bar{b} u_i)= \frac{1}{\Lambda^4} \vec{\alpha}_{bbu_i}\cdot
  \vec{c}_{bbu_i}, \label{eq:master:bbj}
\end{align}
where
\begin{align}
  \vec{c}_{bbu_i}
 \equiv & (|c_{qq}^{1(33i3)}|^2,
|c_{qu}^{1(33i3)}|^2,
|c_{qd}^{1(i333)}|^2,
|c_{ud}^{1(i333)}|^2,
|c_{qq}^{3(33i3)}|^2,
|c_{qu}^{8(33i3)}|^2,
|c_{qd}^{8(i333)}|^2,
|c_{ud}^{8(i333)}|^2,\
\nonumber \\ &
\mathrm{Re}[(c_{qq}^{1(33i3)})(c_{qq}^{3(33i3)})^\ast],
|c_{quqd}^{1(i333)}|^2,
|c_{quqd}^{1(33i3)}|^2,
|c_{quqd}^{1(3i33)}|^2,
|c_{quqd}^{8(i333)}|^2,
|c_{quqd}^{8(33i3)}|^2,
|c_{quqd}^{8(3i33)}|^2,
\nonumber \\ &
\mathrm{Re}[(c_{quqd}^{1(i333)})(c_{quqd}^{1(33i3)})^\ast],
\mathrm{Re}[(c_{quqd}^{8(i333)})(c_{quqd}^{8(33i3)})^\ast],
\mathrm{Re}[(c_{quqd}^{1(i333)})(c_{quqd}^{8(33i3)})^\ast],
\nonumber \\ &
\mathrm{Re}[(c_{quqd}^{8(i333)})(c_{quqd}^{1(33i3)})^\ast],
\mathrm{Re}[(c_{quqd}^{1(3i33)})(c_{quqd}^{8(3i33)})^\ast])^T
  \label{eq:veccbbj}
\end{align}
%
%
%% 
%% %
%
\begin{figure}[t]
 \begin{center}
  \includegraphics[width=0.49\columnwidth]{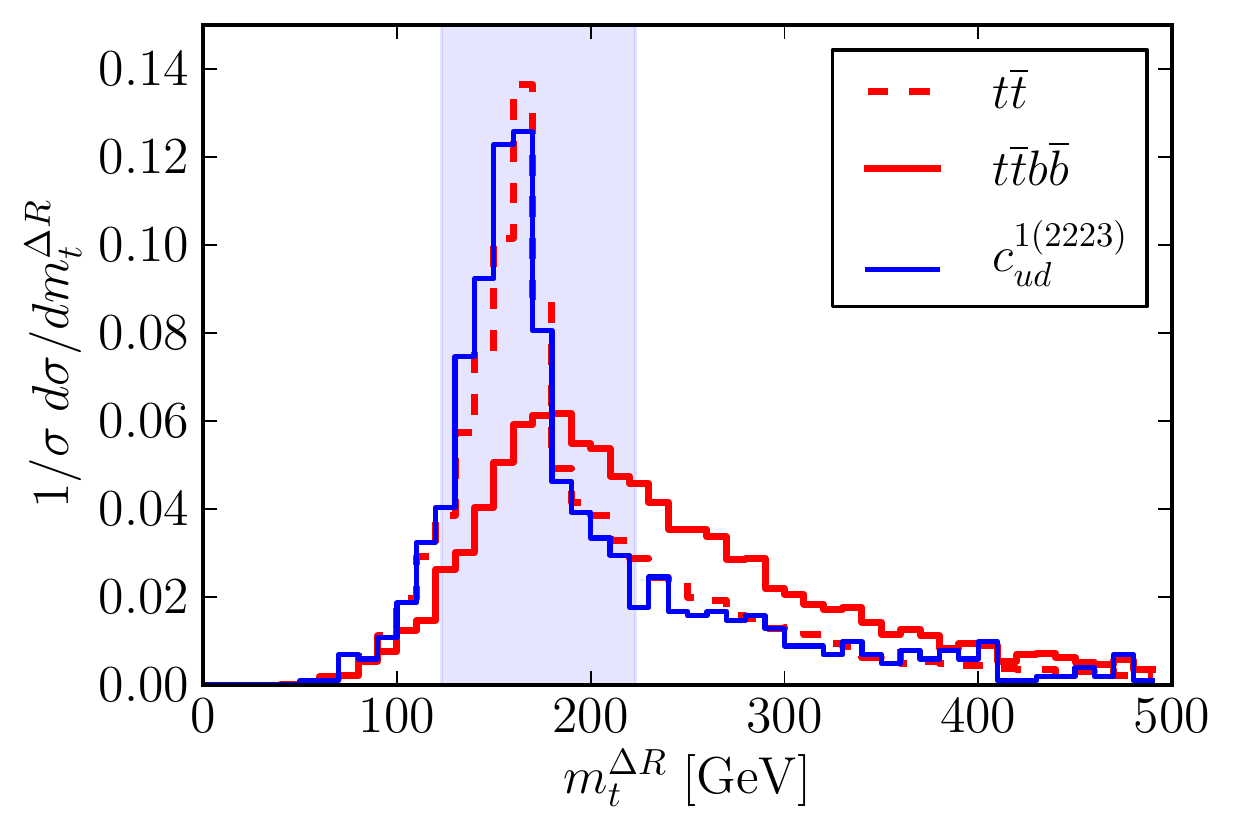}
  \includegraphics[width=0.49\columnwidth]{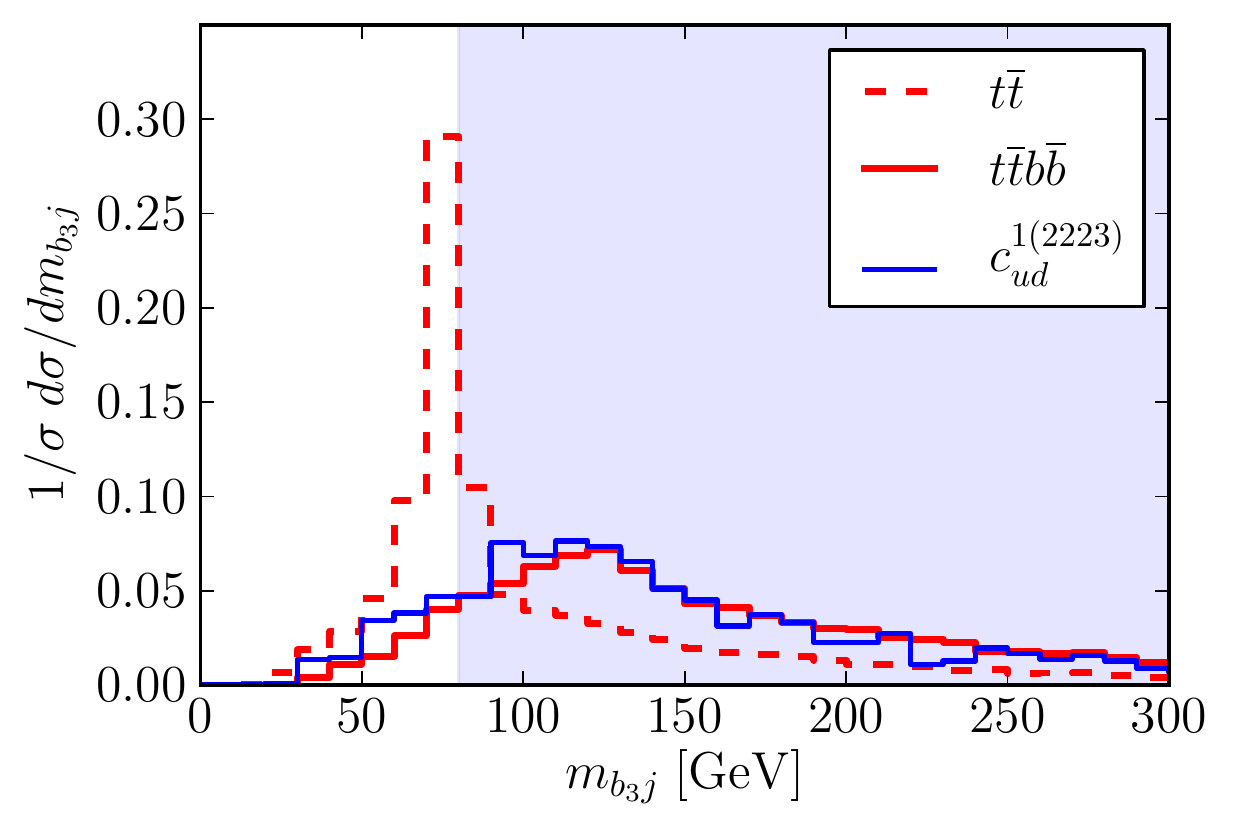}
 \end{center}
\caption{Normalized distributions of $m_t^{\Delta R}$ (left) and $m_{b_3 j}$ (right) in the signal (thin solid blue) and the two main backgrounds: $t\overline{t}$ (thin dashed red) and $t\overline{t}b\overline{b}$ (thick solid red). The cut imposed in our analysis is represented by the blue shaded area.}\label{fig:histo2}
\end{figure}
No experimental analysis tagging the top decay into $b\overline{b}j$ sensitive 
to 
four-fermion operators has been worked out to the date. We demonstrate, however, 
that a sensible reach can be obtained in the HL-LHC.
To this end, we follow closely the analysis of
Ref.~\cite{Banerjee:2018fsx}.  Both muons and electrons are defined by
$p_T^\ell > 10$ GeV and $|\eta_\ell| < 2.5$. Jets are clustered using
the anti-k$_T$ algorithm with $R = 0.4$ and they are required to have
$p_T^j > 30$ GeV and $|\eta_j| < 2.5$. We require the presence of
exactly one isolated lepton and four jets, of which exactly three must
be $b$-tagged. The $b$-tagging efficiency has been fixed to $0.7$; the
charm (light jet) mistag rate being  $0.1$ ($0.01$). We will refer to
this set of cuts as \textit{basic cuts} hereafter. We then obtain the
two $b$-tagged jets closest in $\Delta R$, out of which we reconstruct
the hadronic top mass $m_t^{\Delta R}$ by joining their momenta with
that of the light jet. 
Out of the lepton, the third $b$-jet and the missing energy, we also
construct the transverse mass $m_T$. We require $|m_t^{\Delta R} -
175| < 50$ GeV as well as $m_T < 200$ GeV. Finally, we construct the
invariant mass of the third $b$-tagged jet and the light one, $m_{b_3
  j}$. This usually peaks around the $W$ boson mass in the background;
see Fig.~\ref{fig:histo2}. We thus enforce $m_{b_3 j} > 80$ GeV. The
main background ensues from
\mc{$t\overline{t}$ (including the CKM suppressed $t\rightarrow b W, 
W\rightarrow bc$) merged up to one extra hard jet as well as from  
$t\overline{t}b\overline{b}$, both in the
semi-leptonic (SL) and di-Leptonic (LL) channels. We also} include
the leptonic $Wb\overline{b}$ and $Zb\overline{b}$ merged up to two
extra matrix element partons. For the matching procedure, we
employ the MLM merging scheme~\cite{Mangano:2006rw}. The cross
sections for $t\overline{t}b\overline{b}$, $t\overline{t}$,
$Wb\overline{b}$ and $Zb\overline{b}$ are multiplied by the QCD NLO
$K$-factors of 1.13, 1.6, 2.3 and 1.25,
respectively~\cite{kfactors1,kfactors2}. For both the signal and the
background, we use the \texttt{NNPDF 2.3}~\cite{Ball:2012cx} at
leading order. 

The cutflow is shown in Table~\ref{tab:cutflow}. The efficiency of
selecting \mc{signal events (with one top decaying exotic)} is operator 
independent to very
good accuracy; being roughly $\sim 3.6\times 10^{-3}$. This results in
the coefficients for the master equation, Eq.~(\ref{eq:master:bbj}), shown in
Table~\ref{tab:master:bbj}. The total number of background events is
of order $\sim 3\times 10^5$.
Therefore, \mc{using again the CL$_s$ method we obtain
$s_\mathrm{max}\sim 1.1\times 10^3$ ($6\times 10^4$) under the assumption of 
no systematic uncertainties ($10\,\%$ systematics). This corresponds to
a 95~\% CL exclusion $\mathcal{B}(t\rightarrow 
b\overline{b}j) > 5.9\times 10^{-5}$ ($3\times 10^{-3}$)}
at the HL-LHC.  Using Eq.~(\ref{eq:master:bbj}) with the 
coefficients in
Table~\ref{tab:master:bbj} we obtain, for the particular case of one
operator at a time, 
the bounds in Table~\ref{tab:hadbounds}. The bounds on each operator, computed 
after marginalizing over all operators 
interfering with it, 
become at most 10~\% weaker.
\begin{table}[t]
\centering
%\begin{footnotesize}
\begin{tabular}{| c |	 c  c  c  c  c  c |}
\hline
						  &			 &			&		   &		      &					  &\\[-0.1cm]

Cuts                                              & $t\overline{t}$ (SL) & $t\overline{t}$ (LL) & $Wb\overline{b}$ & $Zb\overline{b}$ & $t\overline{t}b\overline{b}$ (SL) & $t\overline{t}b\overline{b}$ (LL) \\[0.2cm]
\hline
						  &			 &			&		   &		      &					  &\\[-0.1cm]
Basic                                             & 17  	         & 3.9         		& 1.2     	   & 0.44     	      & 220   		                  & 52                   \\[0.2cm]
$|m_{t}^{\Delta R} - m_t| <$ 50 GeV               & 11		         & 1.4            	& 0.38        	   & 0.16              & 100              		  &  17                  \\[0.2cm]
$m_T < 200$ GeV                                   & 8.1		         & 0.75            	& 0.24        	   & 0.12              & 67                    		  &  10                 \\[0.2cm]
$m_{b_3 j} > 80$ GeV                            & 3.0		         & 0.51           	& 0.17        	   & 0.09              & 60                   		  &  7.0                  \\[0.2cm]
\hline
\end{tabular}
%\end{footnotesize}
\caption{\label{tab:sbb2}Cumulated efficiency $\times 10^4$ after each cut for 
the six dominant backgrounds. SL (LL) denotes semi (di)-leptonic decays. 
}\label{tab:cutflow}\vspace{0.5cm}
\end{table}
%%%%%%%%%%%%%%%%%%%%%%%%%%%%%%%
%
\begin{table}[t]
\footnotesize
\begin{center}
 \begin{tabular}{|l|cccccccccccc|}
 \hline
 						  &			 &		
	&		   &		      &					 
 & & & & & & &\\[-0.1cm]
 & $\alpha_{bbu_i}^{1-4}$
 & $\alpha_{bbu_i}^{5}$
 & $\alpha_{bbu_i}^{6-8}$
 & $\alpha_{bbu_i}^{9}$
 & $\alpha_{bbu_i}^{10-11}$
 & $\alpha_{bbu_i}^{12}$
 & $\alpha_{bbu_i}^{13-14}$
 & $\alpha_{bbu_i}^{15}$
 & $\alpha_{bbu_i}^{16}$
 & $\alpha_{bbu_i}^{17}$
 & $\alpha_{bbu_i}^{18-19}$
 & $\alpha_{bbu_i}^{20}$
 \\[0.2cm]
  \hline
  						  &			 &		
	&		   &		      &					 
 & & & & & & &\\[-0.1cm]
  $bbu_j$
  & 1.5
  & 5.4
  & 0.33
  & 0.98
  & 0.37
  & 0.86
  & 0.082
  & 0.14
  & 0.12
  & -0.027
  & 0.16
  & 0.33
  \\[0.2cm]
  \hline
 \end{tabular}
\caption{Coefficients  
\mc{$\times 10^{-2}$} of the master equation (\ref{eq:master:bbj}), in 
$\mathrm{TeV}^4$,
  HL-LHC. See text for the definition of the signal region.}
\label{tab:master:bbj}
\end{center}
\end{table}
\begin{table}[ht]
\centering
%\begin{footnotesize}
%\begin{center}
 \begin{tabular}{| l |c c c c c c c c|}
 \hline
						  &			 &			&		   &		      &					  & & &\\[-0.1cm]
  & $c_{qq}^{1(3323)}$ & $c_{qq}^{3(3323)}$ & $c_{qu}^{1(3323)}$ & $c_{qu}^{8(3323)}$ & $c_{qd}^{1(2333)}$ & $c_{qd}^{8(2333)}$ & $c_{ud}^{1(2333)}$ & $c_{ud}^{8(2333)}$ \\[0.2cm]
  \hline
  						  &			 &			&		   &		      &					  & & &\\[-0.1cm]
 % Bound & 2.9 & 1.5 & 2.9 & 2.9 & 2.9 & 6.2 & 6.2 & 6.2\\[0.2cm]
  Bound & 2.7 & 1.4 & 2.7 & 5.8 & 2.7 & 5.8 & 2.7 & 5.8\\[0.2cm]
  \hline
 \end{tabular}
\end{table}
\begin{table}[ht]
\centering
%\begin{footnotesize}
%\begin{center}
 \begin{tabular}{| l |c c c c c c|}
 \hline
						  &			 &			&		   &		      &					  & \\[-0.1cm]
  & $c_{quqd}^{1(3233)}$ & $c_{quqd}^{1(3323)}$ & $c_{quqd}^{1(2333)}$ & $c_{quqd}^{8(3233)}$ & $c_{quqd}^{8(3323)}$ & $c_{quqd}^{8(2333)}$  \\[0.2cm]
  \hline
  						  &			 &			&		   &		      &					  &\\[-0.1cm]
  Bound & 3.6 & 5.5 & 5.5 & 9.0 & 11.6 & 11.6 \\[0.2cm]
  \hline
 \end{tabular}
% \end{footnotesize}
\caption{Expected bound on $c$ for $\Lambda = 1$ TeV in the HL-LHC 
\mc{assuming no systematics. They become a factor of $\sim 7$ larger for a 
systematic uncertainty of $10\,\%$.}}\label{tab:hadbounds}
%\end{center}
\end{table}

These bounds are competitive with (and often superior to) limits from low energy experiments and other collider searches for top production. 
To start with, the operators above are very poorly constrained by EWPD~\cite{deBlas:2015aea}. The $qq$ operators are severely constrained if all flavour is assumed to be in the down sector. Indeed, in that case the interaction $\sim c_{qq}/\Lambda^2 V_{ts}^{\text{CKM}} (\overline{b_L}\gamma^\mu s_L)(\overline{s_L}\gamma_\mu b_L)$ arises. This interaction is severely constrained by current measurements of the $\overline{B_s}$-$B_s$ mixing parameter $\Delta M_s$ yielding $c_{qq} \lesssim 10^{-4}$ for $\Lambda \sim 1$ TeV~\cite{DiLuzio:2017fdq}. If all flavour is in the up sector, we obtain $\sim c_{qq}/\Lambda^2 V_{ts}^\text{CKM} (V_{td}^\text{CKM})^2 (\overline{u_L}\gamma^\mu c_L)(\overline{u_L}\gamma_\mu c_L)$ and then $D$-$\overline{D}$ mixing excludes $c_{qq}\lesssim 1$; see \textit{e.g.} Ref~\cite{Harnik:2012pb}. (Operators involving RH fields can be instead safe from flavour constraints.) Independently of the flavour assumption, these operators renormalize also dimension-six interactions such as $\mathcal{O}_{qq}^{1(3232)}$, \textit{e.g.}
\begin{equation}
 c_{qq}^{1(3232)} \sim \frac{3 g_2^2}{(4\pi)^2} V^{\text{CKM}}_{ts} c_{qq}^{1(3332)}\log{\frac{v}{\Lambda}} \sim 10^{-4}~c_{qq}^{1(3332)}~,
\end{equation}
with $g_2$ the $SU(2)_L$ gauge coupling.
The bound from $\Delta M_s$ mentioned before
translates however into a negligible limit on the FCNC interactions. Finally, it is also worth mentioning the reach of current measurements of the $B_c^-$ meson lifetime. Bounds on $(\overline{\tau}\gamma^\mu\nu)(\overline{c}\gamma_\mu b)+\text{h.c.}$ four-fermion operators have been obtained \textit{e.g.} in Ref.~\cite{Alonso:2016oyd}. They are $\mathcal{O}(1)$ for a cut-off of $\sim 1$ TeV. However, they are only induced after running from the four-quark operators, being the bounds on the latter therefore too weak.
%

%\

Moreover, these operators contribute in general to single top production in the channel $pp\rightarrow t \overline{b}$ (and anti-particles). To the best of our knowledge, no measurement of the corresponding cross section at $\sqrt{s} = 13$ TeV has been performed. The measurement at $\sqrt{s} = 8$ TeV by the ATLAS and CMS Collaborations can be found in Ref.~\cite{Moles-Valls:2017oed}, yielding $\sigma = 4.8 \pm 0.8 (\text{stat.}) \pm 1.6 (\text{syst.})$ pb. Being already systematically dominated, it will be hard to reduce the uncertainty below the picobarn level. On the other hand, the operators in Table~\ref{tab:hadbounds} contribute to this process via sea quarks in the proton, for which the corresponding cross sections are typically small. For example, setting only $c_{qq}^{3(3323)}/\Lambda^2 = 1$ TeV$^{-2}$, we obtain $\sigma\sim 2$ pb. Therefore, searches for single-top production might improve on our bounds. Making a sharper statement about their reach is however out of the scope of this work.

\section{Beyond the second generation fermions}
\label{sec:beyond}
\begin{figure}[t]
{
 \includegraphics[width = 0.49\columnwidth]{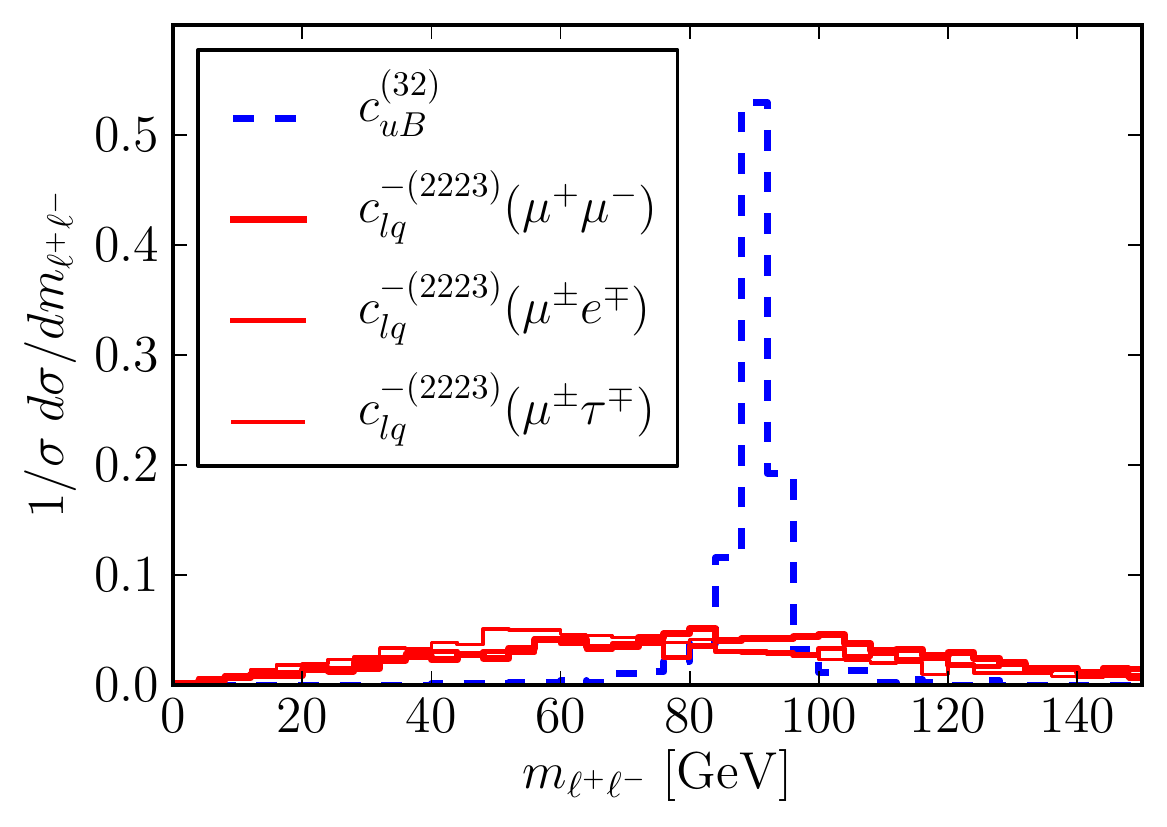}
 \includegraphics[width = 0.49\columnwidth]{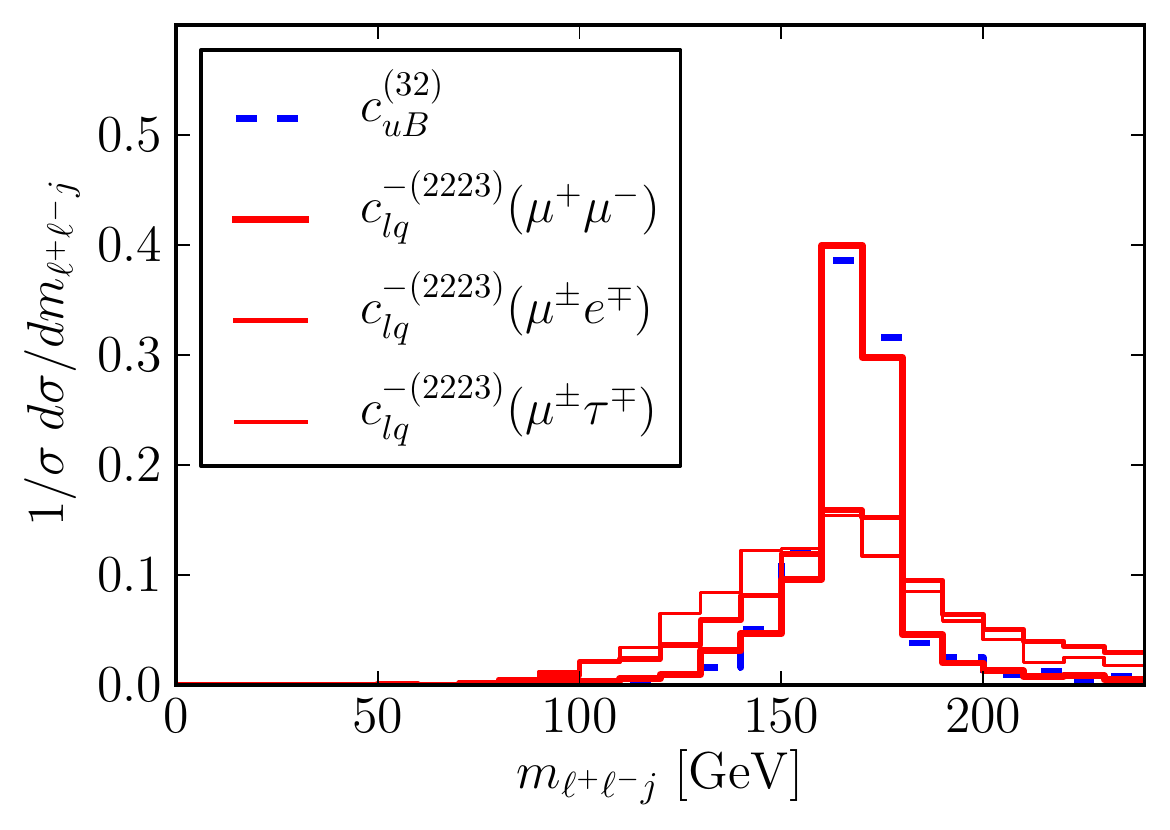}
 \includegraphics[width = 0.49\columnwidth]{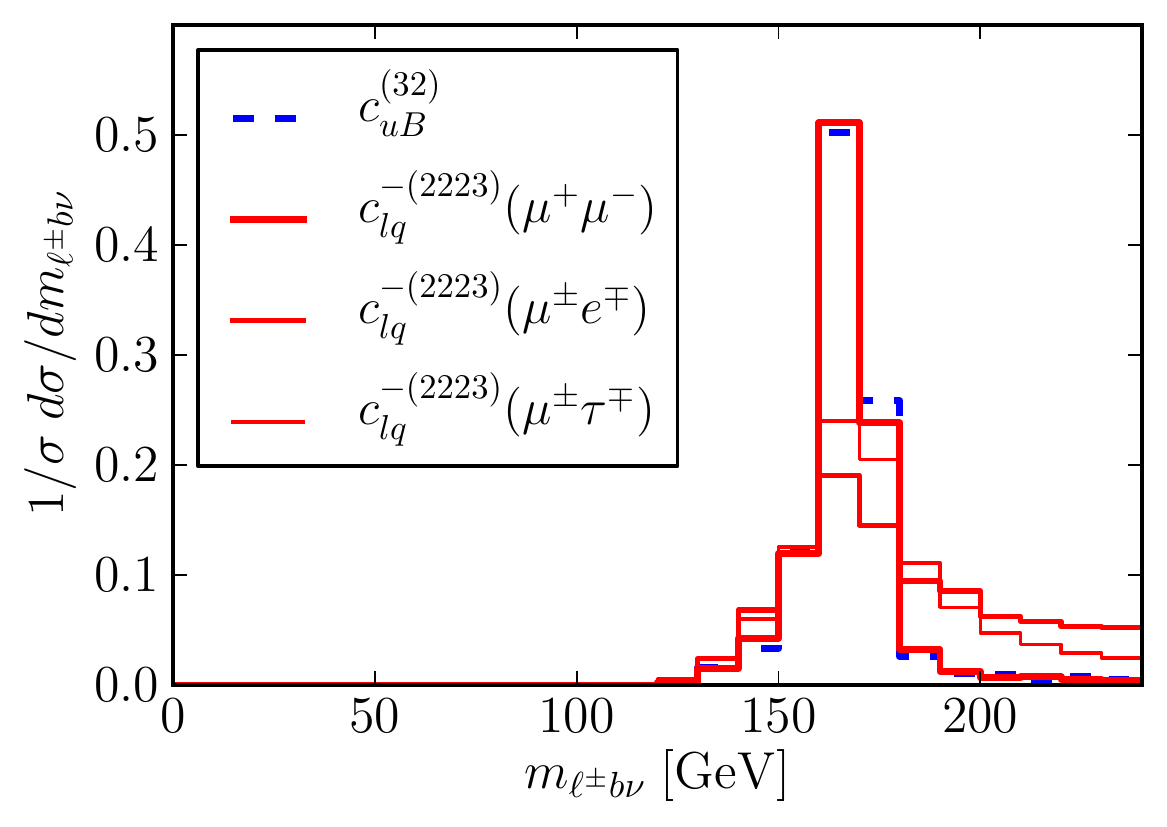}
 \hspace{0.2cm}\includegraphics[width = 0.49\columnwidth]{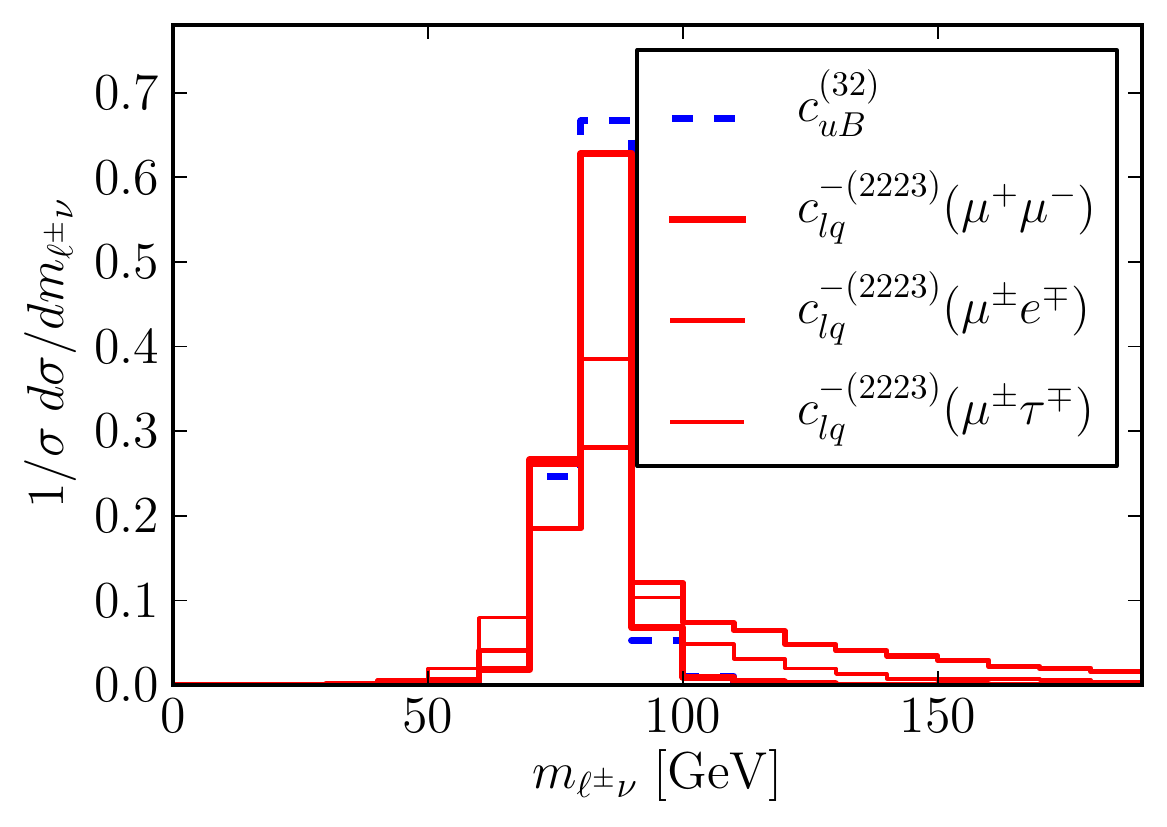}}
 \caption{Top left) Normalized distribution of $m_{\ell^+\ell^-}$ in the three lepton final state of top pair events with one top decaying as $t\to Zj$ (dashed blue), and in the case in which the latter decays as $t\to j \ell^\pm_i\ell_j^\mp$ (solid red) for different LFV combinations. Top right) Same before but for $m_{\ell^+\ell^- j}$. Bottom left) Same as before but for $m_{\ell^\pm b\nu}$. Bottom right) Same as before but for $m_{\ell^\pm \nu}$.}\label{fig:LFVplots}
\end{figure}
So far we have assumed $\ell^\pm = \mu^\pm$, $j = c$. Let us know
relax these assumptions. The main implication of having $j = u$ with
respect to $j = c$ is the smaller misstag rate for $b$-tagging. It
translates into an efficiency for selecting events in this new case of
about $\sim 0.99/0.84\sim 1.18$ larger. (This number agrees perfectly
with the results reported in Ref.~\cite{Aaboud:2018nyl}.) In turn, the
bounds on $c/\Lambda^2$ get around $1.09$ times stronger. Likewise,
two-quark-two-lepton operators might have $\ell^\pm = e^\pm$. The main
difference with respect to the muon case is the smaller efficiency for
reconstructing electrons at the detector level. Being conservative, we
can estimate the reduction in efficiency by a factor of $\sim
(0.77/0.94)^2\sim 0.82^2\sim
0.67$~\cite{daSilvaFernandesdeCastro:2008zz}. In turn, the bounds on
$c/\Lambda^2$ get a factor of $\sim 0.82$ weaker. 

Two-quark-two-lepton interactions might also be LFV. In such case, the
smaller number of events with \mc{SFOS} leptons near the $Z$ pole becomes
far more important than the aforementioned reduced electron
efficiency.  This effect is exacerbated for contact-interactions
involving tau leptons, due to the smaller decay rate of the latter
into light leptons. For LFV operators, not only the distribution of
$m_{\ell^+\ell^-}$ is different with respect to the $t\rightarrow Zj$
case. Also the distributions of $m_{\ell^+\ell^- j}$, $m_{\ell^\pm b
  \nu}$ and $m_{\ell^\pm \nu}$ differ from the latter and among
themselves; see Fig.~\ref{fig:LFVplots}. The overall effect is that
$\mu^\pm e^\mp$, $\mu^\pm\tau^\mp$ and $e^\pm\tau^\mp$ have much
smaller efficiencies in SRA. In CR1, however, the impact is smaller,
being the corresponding efficiencies a factor of $\sim 0.90$, $0.19$
and $0.16$ smaller, respectively. Likewise, the efficiencies in the
new region are reduced by factors of $\sim 0.43$, $0.13$ and $0.11$,
respectively. However, the corresponding top width becomes twice
larger, because the top can decay into $\mu^+ e^-$ and $\mu^- e^+$
(and analogously for taus) instead of just $\mu^+ \mu^-$. Note that
this assumes that both LFV couplings are present.

In summary, the coefficients of the master equation in
Table~\ref{tab:masterlljcoeffs} can be trivially extrapolated to other
final states using the following factors for lepton preserving processes
\begin{align}
  &\alpha(t\to \ell_i^+ \ell_j^- u) = 1.18 \times\alpha(t\to \ell_i^+ \ell_j^-
  c), \nonumber \\
  &\alpha(t\to e^+ e^- j)= 0.67 \times \alpha(t\to \mu^+ \mu^- j).
\end{align}
For LFV processes in the CR1 signal region
\begin{align}
  &\alpha^{CR1}(t\to \mu^\pm e^\mp j)= 1.80 \times
  \alpha^{CR1}(t\to \mu^+ \mu^- j), 
\nonumber \\
&\alpha^{CR1}(t\to \mu^\pm \tau^\mp j)= 0.38 \times\alpha^{CR1}(t\to \mu^+ \mu^- j), 
\nonumber \\
&\alpha^{CR1}(t\to e^\pm \tau^\mp j)= 0.32\times
\alpha^{CR1}(t\to \mu^+ \mu^- j),
\end{align}
and for LFV processes in the NEW signal region
\begin{align}
&\alpha^{NEW}(t\to \mu^\pm e^\mp j)= 0.86 \times\alpha^{NEW}(t\to \mu^+
  \mu^- j), \nonumber \\
&\alpha^{NEW}(t\to \mu^\pm \tau^\mp j)= 0.26 \times\alpha^{NEW}(t\to \mu^+ \mu^- j), \nonumber\\
&\alpha^{NEW}(t\to e^\pm \tau^\mp j)= 0.22 \times
\alpha^{NEW}(t\to \mu^+ \mu^- j).
\end{align}
The current (future) bounds on the four-fermion operators with all
flavours considered in this article in light of the new efficiencies
are given in Tabs.~\ref{tab:finalBounds1} and
\ref{tab:finalBounds2}. They have been obtained using CR1 with the
current luminosity (the new region at the HL-LHC). When obtaining
these numbers we have considered that the two LFV couplings are equal
(for instance $c_{eu}^{1232}=c_{eu}^{2132}$ and the same for all other
operators).
%s
%
%
\afterpage{
\begin{landscape}
\begin{table}[t]
 \centering
 %%%%%%%%%%%%%%%%%%%%%%%%%%%%%%%
 %%%%%%%%%%%%%%%%%%%%%%%%%%%%%%%
 \begin{tabular}{|l|llllllll|}
\hline
 						  &			 &			&		   &		      &					  & & &\\[-0.1cm]
      & $c_{l q}^{-(ij23)}$ & $c_{e q}^{(ij23)}$ & $c_{l u}^{(ij23)}$ & $c_{e u}^{(ij23)}$ & $c_{l equ}^{1(ij23)}$ & $c_{l equ}^{1(ij32)}$ & $c_{l equ}^{3(ij23)}$ & $c_{l equ}^{3(ij32)}$ \\[0.2cm]
\hline
  						  &			 &			&		   &		      &					  & & &\\[-0.1cm]
$\mu^+\mu^-$ & 8.4 (1.0)  & 8.4 (1.0)  & 8.4 (1.0)  & 8.4 (1.0)  & 18.0 (2.2) & 
18.0 (2.2) & 2.3 (0.28) & 2.3 (0.28) \\
      &            &            &            &            &            &         
   &            &            \\
 $\mu^\pm e^\mp$ & 6.3 (1.1)  & 6.3 (1.1)  & 6.3 (1.1)  & 6.3 (1.1)  & 13.0 
(2.4) & 13.0 (2.4) & 1.7 (0.3)  & 1.7 (0.3)  \\
      &            &            &            &            &            &         
   &            &            \\
 $\mu^\pm\tau^\mp$ & 14.0 (2.0) & 14.0 (2.0) & 14.0 (2.0) & 14.0 (2.0) & 29.0 
(4.3) & 29.0 (4.3) & 3.7 (0.55) & 3.7 (0.55) \\
      &            &            &            &            &            &         
   &            &            \\
 $e^+e^-$ & 10.0 (1.2) & 10.0 (1.2) & 10.0 (1.2) & 10.0 (1.2) & 22.0 (2.7) & 
22.0 (2.7) & 2.8 (0.34) & 2.8 (0.34) \\
      &            &            &            &            &            &         
   &            &            \\
 $e^\pm\tau^\mp$ & 15.0 (2.1) & 15.0 (2.1) & 15.0 (2.1) & 15.0 (2.1) & 32.0 
(4.7) & 32.0 (4.7) & 4.1 (0.6)  & 4.1 (0.6)  \\
\hline
\end{tabular}
 %%%%%%%%%%%%%%%%%%%%%%%%%%%%%%%
 %%%%%%%%%%%%%%%%%%%%%%%%%%%%%%%
 \caption{Current (future) bounds on $c$ for $\Lambda=1$ TeV. The
   choice of the lepton flavour indices $i,j$ depend on the
   process $t \to \ell^+_i \ell^-_j c$, as indicated in the first column. In the case of LFV decays
   we assume $c^{ijkl}=c^{jikl}$. \mc{The future bounds become a factor of 
$\sim 3$ weaker for systematic uncertainties on the background of $\sim 10$ 
\%.}}
 \label{tab:finalBounds1}
\end{table}
%%%%%%%%%%%%%%%%%%%%%%%%%%%%%%%%%%
\begin{table}[t]
 \centering
 %%%%%%%%%%%%%%%%%%%%%%%%%%%%%%%
 %%%%%%%%%%%%%%%%%%%%%%%%%%%%%%%
 \begin{tabular}{|l|llllllll|}
\hline
 						  &			 &			&		   &		      &					  & & &\\[-0.1cm]
      & $c_{l q}^{-(ij23)}$ & $c_{e q}^{(ij23)}$ & $c_{l u}^{(ij23)}$ & $c_{e u}^{(ij23)}$ & $c_{l equ}^{1(ij23)}$ & $c_{l equ}^{1(ij32)}$ & $c_{l equ}^{3(ij23)}$ & $c_{l equ}^{3(ij32)}$ \\[0.2cm]
\hline
  						  &			 &			&		   &		      &					  & & &\\[-0.1cm]
% 
%       &            &            &            &            &            &         
%    &            &            \\
% 
$\mu^+\mu^-$ & 7.6 (0.9)  & 7.6 (0.9)  & 7.6 (0.9)  & 7.6 (0.9)  & 16.0 (2.0) & 
16.0 (2.0) & 2.1 (0.25) & 2.1 (0.25) \\
      &            &            &            &            &            &         
   &            &            \\
 $\mu^\pm e^\mp$ & 5.6 (0.97) & 5.6 (0.97) & 5.6 (0.97) & 5.6 (0.97) & 12.0 
(2.1) & 12.0 (2.1) & 1.5 (0.27) & 1.5 (0.27) \\
      &            &            &            &            &            &         
   &            &            \\
 $\mu^\pm\tau^\mp$ & 12.0 (1.8) & 12.0 (1.8) & 12.0 (1.8) & 12.0 (1.8) & 26.0 
(3.9) & 26.0 (3.9) & 3.4 (0.49) & 3.4 (0.49) \\
      &            &            &            &            &            &         
   &            &            \\
 $e^+ e^-$ & 9.2 (1.1)  & 9.2 (1.1)  & 9.2 (1.1)  & 9.2 (1.1)  & 20.0 (2.4) & 
20.0 (2.4) & 2.5 (0.31) & 2.5 (0.31) \\
      &            &            &            &            &            &         
   &            &            \\
 $e^\pm\tau^\mp$ & 13.0 (1.9) & 13.0 (1.9) & 13.0 (1.9) & 13.0 (1.9) & 29.0 
(4.2) & 29.0 (4.2) & 3.7 (0.54) & 3.7 (0.54) \\
\hline
\end{tabular}
 %%%%%%%%%%%%%%%%%%%%%%%%%%%%%%%
 %%%%%%%%%%%%%%%%%%%%%%%%%%%%%%%
 \caption{Same as Table~\ref{tab:finalBounds1} for the process
   $t\to \ell^+_i \ell^-_j u$. }\label{tab:finalBounds2}
\end{table}
\end{landscape}
}

Obviously, LFV interactions are also constrained by low-energy experiments. However, the latter are not necessarily better than the LHC. We note, for example, that $c_{lq}^{1223)}/\Lambda^2$ can be bounded to $\sim 1.1$ TeV$^{-2}$ in the $\mu^\pm e^\mp$ channel. \textit{A priori}, such interaction contributes also to the process $\mu\rightarrow e\gamma$ upon closing the quark loop. At low energies, this process is mediated by the $U(1)_Q$ invariant operator $\sim \overline{\mu_L}\sigma_{\mu\nu}e_R F^{\mu\nu}+\text{h.c.}$ The latter arises from the  dimension-six gauge invariant operators $\overline{L}\sigma_{\mu\nu} H e_R B^{\mu\nu}+\text{h.c.}$ and $\overline{L}\sigma_{\mu\nu} \sigma^i H e_R W_i^{\mu\nu}+\text{h.c.}$ Its size can be fairly estimated from the RGE mixing of $\mathcal{O}_{\ell q}$ with the latter two. Interestingly, it vanishes at one loop. If it arises at two loops, we obtain
\begin{equation}
 \dfrac{\Gamma(\mu\rightarrow e\gamma)}{\Gamma_\mu} \sim \frac{y_\mu^2(V^{\text{CKM}}_{cb})^2}{(4\pi)^8}\frac{m_\mu^3 v^2/\Lambda^4}{m_\mu^5/m_W^4} \sim \frac{g_2^4  (V^{\text{CKM}}_{cb})^2}{(4\pi)^8} \frac{v^4}{\Lambda^4}\sim 2\times 10^{-15}~,
\end{equation}
for $\Lambda\sim 1$ TeV$^{-2}$. (We have approximated the muon width $\Gamma_\mu$ by the SM value $\sim m_\mu^5/m_W^4$; $y_\mu$ stands for the muon Yukawa.) This value is two orders of magnitude smaller than the current best bound, namely $\mathcal{B}(\mu\rightarrow e\gamma) < 4.2\times 10^{-13}$ at the 90 \% CL~\cite{Tanabashi:2018abc}.  Moreover, the computation above is equally valid for $\tau\rightarrow \mu(e)\gamma$, for which the HL-LHC can still provide bounds much stronger than current limits from low-energy experiments, $\mathcal{B}(\tau\rightarrow \mu(e)\gamma) < 4.4 (3.3)\times 10^{-8}$ at the 90 \% CL~\cite{Tanabashi:2018abc}. RR operators are further suppressed, because the $W$ boson does not couple to RH currents.
%
%\
Similar results apply to the other two-quark-two-lepton operators, with the exception of $\mathcal{O}_{lequ}$, which do renormalize the operators contributing to $\mu(\tau)\rightarrow e(\mu)\gamma$ at one loop, being therefore tightly constrained.

\section{Matching UV models}
\label{sec:matching}
In renormalizable weakly-coupled UV completions of the SMEFT, only new
scalars and vectors can generate the operators in
Eqs.~\ref{eq:lepop1}-\ref{eq:hadop2} upon integrating out at tree level. The only
scalars are: $\omega_1\sim (3,1)_{-\frac{1}{3}}$, $\zeta\sim
(3,3)_{-\frac{1}{3}}$, $\Pi_7\sim (3,2)_{\frac{7}{6}}$, $\varphi\sim
(1,2)_{\frac{1}{2}}$, $\Omega_1\sim (6,1)_{\frac{1}{3}}$,
$\Upsilon\sim (6,3)_{\frac{1}{3}}$ and $\Phi\sim
(8,2)_{\frac{1}{2}}$. The numbers in parentheses and the subscript
indicate their representations under $SU(3)_c$ and $SU(2)_L$ and the
hypercharge, respectively. The relevant couplings of these particles
read: 
\begin{align}\label{eq:scalars1}
  L_{\omega_1}&=-\bigg\lbrace
  (y^{ql}_{\omega_1})_{rij} \omega_{1r}^\dagger
    \bar{q}^c_{Li} \mathrm{i} \sigma_2 l_{Lj}
   +(y^{eu}_{\omega_1})_{rij} \omega_{1r}^\dagger
   \bar{e}^c_{Ri}  u_{Rj} \\
   &\quad+ (y^{qq}_{\omega_1})_{rij} \omega_{1r}^\dagger
   \epsilon_{ABC}  \bar{q}^B_{Li} \mathrm{i} \sigma_2 q^{cC}_{Lj}
+  (y^{du}_{\omega_1})_{rij} \omega_{1r}^\dagger
   \epsilon_{ABC}  \bar{d}^B_{Ri} u^{cC}_{Rj}+ \mathrm{h.c.}\bigg\rbrace + \ldots~,\\
  L_{\zeta}&=-\bigg\lbrace
  (y^{ql}_{\zeta})_{rij} \zeta_{r}^{a\,\dagger}
    \bar{q}^c_{Li} \mathrm{i} \sigma_2 \sigma^a l_{Lj}+(y^{qq}_{\zeta})_{rij} \zeta_{r}^{a\,\dagger}
    \epsilon_{ABC} \bar{q}^B_{Li} \sigma^a \mathrm{i} \sigma_2  q^{cC}_{Lj}
 + \mathrm{h.c.}\bigg\rbrace + \ldots~,\\
%
%      \end{align}
%      \begin{align}\label{eq:scalars2}
      %
  L_{\Pi_7}&=-\bigg\lbrace
  (y^{lu}_{\Pi_7})_{rij} \Pi_{7r}^\dagger
    \mathrm{i} \sigma_2  \bar{l}^T_{Li} u_{Rj}
+  (y^{eq}_{\Pi_7})_{rij} \Pi_{7r}^\dagger
    \bar{e}_{Ri} q_{Lj}
      + \mathrm{h.c.}\bigg\rbrace~,\\
  L_{\varphi}&=-\bigg\lbrace
  (y^{e}_{\varphi})_{rij} \varphi_{r}^\dagger \bar{e}_{Ri} l_{Lj}
+(y^u_{\varphi})_{rij} \varphi_r^\dagger  \mathrm{i} \sigma_2  \bar{q}^T_{Li} u_{Rj}
      +(y^{d}_{\varphi})_{rij} \varphi_{r}^\dagger \bar{d}_{Ri} q_{Lj}+ \mathrm{h.c.}\bigg\rbrace + \ldots~,\\
 L_{\Omega_1}&=-\bigg\lbrace
  (y^{ud}_{\Omega_1})_{rij} \Omega_{1r}^{AB\,\dagger}
  \bar{u}_{Ri}^{c(A|} d^{|B)}_{Rj}
+ (y^{qq}_{\Omega_1})_{rij} \Omega_{1r}^{AB\,\dagger}
  \bar{q}_{Li}^{c(A|} \mathrm{i} \sigma_2 q^{|B)}_{Lj}
 + \mathrm{h.c.}\bigg\rbrace~,\\
 L_{\Upsilon} &=-\left\lbrace
  (y_{\Upsilon})_{rij} \Upsilon_{r}^{aAB\,\dagger}
  \bar{q}_{Li}^{c(A|} \mathrm{i} \sigma_2 \sigma^a q^{|B)}_{Lj}
 + \mathrm{h.c.}\right\rbrace~,\\
  L_{\Phi}&=-\left\lbrace
  (y_{\Phi}^{qu})_{rij} \Phi_{r}^{A\dagger}
  \mathrm{i}\sigma_2 \bar{q}_{Li}^{T} T_A u_{Rj}
+  (y_{\Phi}^{dq})_{rij} \Phi_{r}^{A\dagger}
  \bar{d}_{Ri} T_A q_{Lj}
 + \mathrm{h.c.}\right\rbrace~.
 \end{align}
The ellipsis indicate that other couplings are in general present, but they do not contribute to the operators under study. ($\Pi_7$, therefore, does not induce any other operator.) Moreover, only in the case of $\varphi$, the couplings explicitly shown in Eqs.~\ref{eq:scalars1} generate also interactions not considered in this article; see Refs.~\cite{deBlas:2014mba,deBlas:2017xtg} for details.

\afterpage{\begin{landscape}
\begin{table}[t]
\centering
%\begin{footnotesize}
%\begin{center}
 \begin{tabular}{| l |c c c c c c|}
 \hline
						  &			 &			&		   &						  & &\\[-0.1cm]
  & $c_{lq}^{-(ijkl)}$ & $c_{eq}^{(ijlk)}$ & $c_{lu}^{(ijkl)}$ & $c_{eu}^{(ijkl)}$ & $c_{lequ}^{1(ijkl)}$ & $c_{lequ}^{3(ijkl)}$  \\[0.4cm]
  \hline
  						  &			 &			&		   &    					  & &\\[-0.1cm]
$\omega_1$ &	$\frac{(y^{ql}_{\omega_1})_{ki}^\ast(y^{ql}_{\omega_1})_{lj}}{2M^2_{\omega_{1}}}$	&			&		&	$\frac{(y^{eu}_{\omega_1})_{ik}^\ast(y^{eu}_{\omega_1})_{jl}}{2M^2_{\omega_{1}}}$	&	$\frac{(y^{eu}_{\omega_1})_{jl}(y^{ql}_{\omega_1})^\ast_{ki}}{2M^2_{\omega_{1}}}$		& $-\frac{(y^{eu}_{\omega_1})_{jl}(y^{ql}_{\omega_1})^\ast_{ki}}{8M^2_{\omega_{1}}}$ \\[0.4cm]
$\zeta$ &	$\frac{(y^{ql}_\zeta)_{ki}^\ast(y^{ql}_\zeta)_{lj}}{2M^2_{\zeta}}$	&			&		&		&			&  \\[0.4cm]
$\Pi_7$ &		&	$-\frac{(y^{eq}_{\Pi_7})_{jk}^\ast(y^{eq}_{\Pi_7})_{il}}{2M^2_{\Pi_{7}}}$		&	$-\frac{(y^{lu}_{\Pi_7})_{jk}^\ast(y^{lu}_{\Pi_7})_{il}}{2M^2_{\Pi_{7}}}$	&		&		$\frac{(y^{eq}_{\Pi_7})_{jk}^\ast(y^{lu}_{\Pi_7})_{il}}{2M^2_{\Pi_{7}}}$	& $\frac{(y^{eq}_{\Pi_7})_{jk}^\ast(y^{lu}_{\Pi_7})_{il}}{8M^2_{\Pi_{7}}}$ \\[0.4cm]
$\varphi$ &		&			&		&		&	$\frac{(y^{u}_{\varphi})_{kl} (y^{e}_{\varphi})_{ij}^\ast}{M^2_{\varphi}}$		&  \\[0.4cm]
$\mathcal{B}$ &	$-\frac{(g^q_{\mathcal{B}})_{kl}(g^l_{\mathcal{B}})_{ij}}{M^2_{\mathcal{B}}}$	& $-\frac{(g^e_{\mathcal{B}})_{kl}(g^q_{\mathcal{B}})_{ij}}{M^2_{\mathcal{B}}}$	$-\frac{(g^u_{\mathcal{B}})_{kl}(g^l_{\mathcal{B}})_{ij}}{M^2_{\mathcal{B}}}$		&		&	$ -\frac{(g^u_{\mathcal{B}})_{kl}(g^e_{\mathcal{B}})_{ij}}{M^2_{\mathcal{B}}}$	&			&  \\[0.4cm]
$\mathcal{W}$ &	$\frac{(g^{q}_{\mathcal{W}})_{kl}(g^{l}_{\mathcal{W}})_{ij}}
    {4M^2_{\mathcal{W}}} $	&			&		&		&			&  \\[0.4cm]
$\mathcal{Q}_1$ &		&			&	$-\frac{(g_{\mathcal{Q}_1}^{ul})^*_{ki}(g_{\mathcal{Q}_1}^{ul})_{lj}}
    {M^2_{\mathcal{Q}_1}} $	&		&			&  \\[0.4cm]
$\mathcal{Q}_5$ &		&		$\frac{(g_{\mathcal{Q}_5}^{eq})^*_{ik}(g_{\mathcal{Q}_5}^{eq})_{jl}}
    {M^2_{\mathcal{Q}_5}}$	&		&		&			&  \\[0.4cm]
  \hline
 \end{tabular}
% \end{footnotesize}
\caption{Field-by-field contribution to two-quark-two-lepton operators. }\label{tab:UVmodels1}
%\end{center}
\end{table}
\end{landscape}}

\afterpage{\begin{landscape}
\begin{table}[t]
\tiny
\centering
%\begin{footnotesize}
%\begin{center}
 \begin{tabular}{| l |c c c c c c c c|}
 \hline
						  &			 &			&		   &						  & & & &\\[-0.2cm]
						  & $c_{qq}^{1(ijkl)}$ & $c_{qq}^{3(ijlk)}$ & $c_{qu}^{1(ijkl)}$ & $c_{qu}^{8(ijkl)}$ & $c_{qd}^{1(ijkl)}$ & $c_{qd}^{8(ijkl)}$ & $c_{ud}^{1(ijkl)}$ & $c_{ud}^{8(ijkl)}$\\[0.2cm]
						  \hline
						    						  &			 &			&		   &    					  & &&&\\[-0.1cm]
$\omega_1$ &	$\frac{(y^{qq}_{\omega_1})_{ik}(y^{qq}_{\omega_1})_{lj}^\ast}{2M^2_{\omega_{1}}}$	&		$- \frac{(y^{qq}_{\omega_1})_{ki}(y^{qq}_{\omega_1})_{jl}^\ast}{2M^2_{\omega_{1}}}$	&		&		&		&	& $\frac{(y^{du}_{\omega_1})_{lj}^\ast(y^{du}_{\omega_1})_{ki}}{3M^2_{\omega_{1}}}$ & $-\frac{(y^{du}_{\omega_1})_{lj}^\ast(y^{du}_{\omega_1})_{ki}}{M^2_{\omega_{1}}}$ \\[0.2cm]
$\zeta$ &	$\frac{3(y^{qq}_\zeta)_{ki}(y^{qq}_\zeta)_{lj}^\ast}{2M^2_{\zeta}}$	&	$-\frac{(y^{qq}_\zeta)_{ki}(y^{qq}_\zeta)_{jl}^\ast}{2M^2_{\zeta}}$		&		&		&			&  & &\\[0.2cm]
$\varphi$ &		&			&	$-\frac{(y^{u}_{\varphi})_{jk}^\ast (y^{u}_{\varphi})_{il}}{6M^2_{\varphi}}$	&	$-\frac{(y^{u}_{\varphi})_{jk}^\ast (y^{u}_{\varphi})_{il}}{M^2_{\varphi}}$	&		$-\frac{(y^{d}_{\varphi})_{li}^\ast (y^{d}_{\varphi})_{kj}}{6M^2_{\varphi}}$	& $-\frac{(y^{d}_{\varphi})_{li}^\ast (y^{d}_{\varphi})_{kj}}{M^2_{\varphi}}$ & &\\[0.4cm]
$\Omega_1$ &	$\frac{(y^{qq}_{\Omega_1})_{ik}^\ast(y^{qq}_{\Omega_1})_{jl}}{4M^2_{\Omega_{1}}}$	&		$\frac{(y^{qq}_{\Omega_1})_{ik}^\ast(y^{qq}_{\Omega_1})_{lj}}{4M^2_{\Omega_{1}}}$	&		&		&			&  & $\frac{(y^{ud}_{\Omega_1})_{ik}^\ast(y^{ud}_{\Omega_1})_{jl}}{3M^2_{\Omega_{1}}}$ & $\frac{(y^{ud}_{\Omega_1})_{ik}^\ast(y^{ud}_{\Omega_1})_{jl}}{2M^2_{\Omega_{1}}}$ \\[0.2cm]
$\Upsilon$ &	$\frac{3(y_{\Upsilon})_{lj}(y_{\Upsilon})_{ki}^*}{4M^2_{\Upsilon}}$	&	$\frac{(y_{\Upsilon})^*_{ki}(y_{\Upsilon})_{jl}}{4M^2_{\Upsilon}}$		&		&		&			&  & &\\[0.2cm]
$\Phi$ &		&			&	$-\frac{2(y^{qu}_{\Phi})_{jk}^*(y^{qu}_{\Phi})_{il}}{9M^2_{\Phi}}$	&	$\frac{(y^{qu}_{\Phi})_{jk}^*(y^{qu}_{\Phi})_{il}}{6M^2_{\Phi}}$	&	$-\frac{2(y^{dq}_{\Phi})_{li}^*(y^{dq}_{\Phi})_{kj}}{9M^2_{\Phi}}$		& $\frac{(y^{dq}_{\Phi})_{li}^*(y^{dq}_{\Phi})_{kj}}{6M^2_{\Phi}}$ & &\\[0.2cm]
$\mathcal{B}$ &	$-\frac{(g^q_{\mathcal{B}})_{kl}(g^q_{\mathcal{B}})_{ij}}{2M^2_{\mathcal{B}}}$	&		&	$-\frac{(g^u_{\mathcal{B}})_{kl}(g^q_{\mathcal{B}})_{ij}}{M^2_{\mathcal{B}}}$	&		&		$-\frac{(g^d_{\mathcal{B}})_{kl}(g^q_{\mathcal{B}})_{ij}}{M^2_{\mathcal{B}}}$	& & $-\frac{(g^u_{\mathcal{B}})_{ij}(g^d_{\mathcal{B}})_{kl}}{M^2_{\mathcal{B}}}$ & \\[0.2cm]
$\mathcal{B}_1$ &		&		&		&		&			& & $-\frac{(g^{du}_{\mathcal{B}_1})^*_{li}(g^{du}_{\mathcal{B}_1})_{kj}}
    {3M^2_{\mathcal{B}_1}}$ & $-\frac{2(g^{du}_{\mathcal{B}_1})^*_{li}(g^{du}_{\mathcal{B}_1})_{kj}}
    {M^2_{\mathcal{B}_1}}$ \\[0.2cm]
$\mathcal{W}$ &		&	$-\frac{(g^{q}_{\mathcal{W}})_{kl}(g^{q}_{\mathcal{W}})_{ij}}
    {8M^2_{\mathcal{W}}}$	&		&		&			& && \\[0.2cm]
$\mathcal{G}$ &	$-\frac{(g^{q}_{\mathcal{G}})_{kj}(g^{q}_{\mathcal{G}})_{il}}
    {8M^2_{\mathcal{G}}}$	&	$-\frac{(g^{q}_{\mathcal{G}})_{kj}(g^{q}_{\mathcal{G}})_{il}}
    {8M^2_{\mathcal{G}}}$	&		&	$-\frac{(g^{u}_{\mathcal{G}})_{kl}(g^{q}_{\mathcal{G}})_{ij}}
    {M^2_{\mathcal{G}}}$	&			& $-\frac{(g^{d}_{\mathcal{G}})_{kl}(g^{q}_{\mathcal{G}})_{ij}}
    {M^2_{\mathcal{G}}} $ & & $-\frac{(g^{d}_{\mathcal{G}})_{kl}(g^{u}_{\mathcal{G}})_{ij}}
    {M^2_{\mathcal{G}}}$ \\[0.2cm]
$\mathcal{G}_1$ &		&		&		&		&			&  & $-\frac{4(g_{\mathcal{G}_1})_{li}^*(g_{\mathcal{G}_1})_{kj}}
    {9M^2_{\mathcal{G}_1}} $ & $\frac{(g_{\mathcal{G}_1})_{li}^*(g_{\mathcal{G}_1})_{kj}}
    {3M^2_{\mathcal{G}_1}} $ \\[0.2cm]
$\mathcal{H}$ &	$\frac{3(g_{\mathcal{H}})_{kj}(g_{\mathcal{H}})_{il}}
    {32M^2_{\mathcal{H}}} $	&	$\frac{(g_{\mathcal{H}})_{kl}(g_{\mathcal{H}})_{ij}}
    {48M^2_{\mathcal{H}}} 
    +\frac{(g_{\mathcal{H}})_{kj}(g_{\mathcal{H}})_{il}}
    {32M^2_{\mathcal{H}}} $	&		&		&			&  &&\\[0.2cm]
$\mathcal{Q}_1$ &		&		&		&		&	$\frac{2(g_{\mathcal{Q}_1}^{dq})^*_{lj}(g_{\mathcal{Q}_1}^{dq})_{ki}}
    {3M^2_{\mathcal{Q}_1}}$		& $-\frac{2(g_{\mathcal{Q}_1}^{dq})^*_{lj}(g_{\mathcal{Q}_1}^{dq})_{ki}}
    {M^2_{\mathcal{Q}_1}} $ &&\\[0.2cm]
$\mathcal{Q}_5$ &		&		&	$\frac{2(g_{\mathcal{Q}_5}^{uq})^*_{lj}(g_{\mathcal{Q}_5}^{uq})_{ki}}
    {3M^2_{\mathcal{Q}_5}}$	&	$- \frac{2(g_{\mathcal{Q}_5}^{uq})^*_{lj}(g_{\mathcal{Q}_5}^{uq})_{ki}}
    {M^2_{\mathcal{Q}_5}}$	&			&  &&\\[0.2cm]
$\mathcal{Y}_1$ &		&		&		&		&	$\frac{2(g_{\mathcal{Y}_1})^*_{lj}(g_{\mathcal{Y}_1})_{ki}}
    {3M^2_{\mathcal{Y}_1}}$		& $\frac{(g_{\mathcal{Y}_1})^*_{lj}(g_{\mathcal{Y}_1})_{ki}}
    {M^2_{\mathcal{Y}_1}}$ & & \\[0.2cm]
$\mathcal{Y}_5$ &		&		&	$\frac{2(g_{\mathcal{Y}_5})^*_{lj}(g_{\mathcal{Y}_5})_{ki}}
    {3M^2_{\mathcal{Y}_5}} $	&	$\frac{(g_{\mathcal{Y}_5})^*_{lj}(g_{\mathcal{Y}_5})_{ki}}
    {M^2_{\mathcal{Y}_5}} $	&			&  &&\\[0.2cm]
 \hline
 \end{tabular}
% \end{footnotesize}
\caption{Field-by-field contribution to four-quark operators.}\label{tab:UVmodels2}
%\end{center}
\end{table}
\end{landscape}
}

On the vector side, the only possible additions are: $ \mathcal{B}\sim (1,1)_0$, $ \mathcal{B}_1\sim (1,1)_1$, $ \mathcal{W}\sim (1,3)_0$, $ \mathcal{G}\sim (8,1)_0$, $ \mathcal{G}_1\sim (8,1)_1$, $ \mathcal{H}\sim (8,3)_0$,  $ \mathcal{Q}_1\sim (3,2)_{\frac{1}{6}}$, $ \mathcal{Q}_5\sim (3,2)_{-\frac{5}{6}}$, $\mathcal{Y}_1\sim (\bar{6},2)_{\frac{1}{6}}$ and $\mathcal{Y}_5\sim (\bar{6},2)_{-\frac{5}{6}}$. Their relevant couplings read:
\begin{align}\label{eq:vectors2}
 L_{\mathcal{B}}&=- \mathcal{B}^\mu \bigg\lbrace
  (g^l_\mathcal{B})_{ij} \bar{l}_{Li} \gamma_\mu l_{Lj}
+  (g^q_\mathcal{B})_{ij} \bar{q}_{Li} \gamma_\mu q_{Lj}
+  (g^e_\mathcal{B})_{ij} \bar{e}_{Ri} \gamma_\mu e_{Rj}
+  (g^u_\mathcal{B})_{ij} \bar{u}_{Ri} \gamma_\mu u_{Rj}\\
&\qquad+  (g^d_\mathcal{B})_{ij} \bar{d}_{Ri} \gamma_\mu d_{Rj}~
\bigg\rbrace+\ldots~,\\
L_{\mathcal{B}_1} &=\left\{\mathcal{B}_1^{\mu\,\dagger} 
  (g^{du}_{\mathcal{B}_1})_{ij} \bar{d}_{Ri} \gamma_\mu u_{Rj}
  +\mathrm{h.c.}
\right\}+\ldots~,\\
 \end{align}
\begin{align}\label{eq:vectors3}
L_{\mathcal{W}} &=- \mathcal{W}^{\mu\,a} \bigg\lbrace
 \frac{1}{2} (g^{l}_{\mathcal{W}})_{ij} \bar{l}_{Li} \sigma^a \gamma_\mu l_{Lj}
+  \frac{1}{2} (g^{q}_{\mathcal{W}})_{ij} \bar{q}_{Li} \sigma^a \gamma_\mu q_{Lj}
\bigg\rbrace+\ldots~,\\
L_{\mathcal{G}} &= \mathcal{G}^{\mu\,A} \bigg\lbrace
 (g^{q}_{\mathcal{G}})_{ij} \bar{q}_{Li} T^A \gamma_\mu q_{Lj}
  + (g^{u}_{\mathcal{G}})_{ij} \bar{u}_{Ri} T^A \gamma_\mu u_{Rj}
  + (g^{d}_{\mathcal{G}})_{ij} \bar{d}_{Ri} T^A \gamma_\mu d_{Rj}
\bigg\rbrace+\ldots~,\\
L_{\mathcal{G}_1} &=-\left\{ \mathcal{G}^{\mu\,A\,\dagger}_1 
 (g_{\mathcal{G}_1})_{ij} \bar{d}_{Ri} T^A \gamma_\mu u_{Rj}
+\mathrm{h.c.}
\right\}~,\\
L_{\mathcal{H}} &= -\frac{1}{2}\left\{ \mathcal{H}^{\mu\,aA} 
 (g_{\mathcal{H}})_{ij} \bar{q}_{Li}\sigma^a T^A \gamma_\mu q_{Lj}
\right\}~,\\
L_{\mathcal{Q}_1}&=-\bigg\lbrace
  \mathcal{Q}_1^{\mu\,\dagger} 
 (g^{ul}_{\mathcal{Q}_1})_{ij} \bar{u}^c_{Ri}\gamma_\mu l_{Lj}
+  \mathcal{Q}_1^{A\,\mu\,\dagger} \epsilon_{ABC}
(g^{dq}_{\mathcal{Q}_1})_{ij} \bar{d}^B_{Ri}\gamma_\mu \mathrm{i}
\sigma_2 q^{cC}_{Lj}
+\mathrm{h.c.}
\bigg\rbrace~,\\
L_{\mathcal{Q}_5}&=-\bigg\lbrace
  \mathcal{Q}_5^{\mu\,\dagger} 
 (g^{dl}_{\mathcal{Q}_5})_{ij} \bar{d}^c_{Ri}\gamma_\mu l_{Lj}
+  \mathcal{Q}_5^{\mu\,\dagger} 
 (g^{eq}_{\mathcal{Q}_5})_{ij} \bar{e}^c_{Ri}\gamma_\mu q_{Lj}\\
&\quad+  \mathcal{Q}_5^{A\,\mu\,\dagger} \epsilon_{ABC}
(g^{uq}_{\mathcal{Q}_5})_{ij} \bar{u}^B_{Ri}\gamma_\mu 
 q^{cC}_{Lj}
+\mathrm{h.c.}
\bigg\rbrace~,\\
 \mathcal{Y}_1 &= -\left\{
  \frac{1}{2} (g_{\mathcal{Y}_1})_{ij}
  \mathcal{Y}_1^{AB\mu\,\dagger} 
  \bar{d}^{(A|}_{Ri}\gamma_\mu \mathrm{i}\sigma_2 q^{c|B)}_{Lj}
+\mathrm{h.c.}
\right\}~,\\
\mathcal{Y}_5 &= -\left\{
  \frac{1}{2} (g_{\mathcal{Y}_5})_{ij}
  \mathcal{Y}_5^{AB\mu\,\dagger} 
  \bar{u}^{(A|}_{Ri}\gamma_\mu \mathrm{i}\sigma_2 q^{c|B)}_{Lj}
+\mathrm{h.c.}
\right\}~.
\end{align}
\begin{table}[t]
\centering
%\begin{footnotesize}
%\begin{center}
 \begin{tabular}{| l |c c|}
 \hline
						  &			 &			\\[-0.1cm]
  & $c_{quqd}^{1((ijkl)}$ & $c_{quqd}^{8(ijlk)}$  \\[0.4cm]
  \hline
  						  &			 &\\[-0.1cm]
$\omega_1$ &	$\dfrac{4(y^{qq}_{\omega_1})_{ki}(y^{du}_{\omega_1})_{lj}^\ast}{3M^2_{\omega_{1}}}$	&	$-\dfrac{4(y^{qq}_{\omega_1})_{ki}(y^{du}_{\omega_1})_{lj}^\ast}{M^2_{\omega_{1}}}$	 \\[0.4cm]
$\varphi$ &	$-\dfrac{(y^{u}_{\varphi})_{ij} (y^{d}_{\varphi})_{lk}^\ast}{M^2_{\varphi}}$	&		 \\[0.4cm]
$\Omega_1$ &	$\dfrac{4(y^{qq}_{\Omega_1})_{ki}^\ast(y^{ud}_{\Omega_1})_{jl}}{3M^2_{\Omega_{1}}}$	&	$\dfrac{2(y^{qq}_{\Omega_1})_{ki}^\ast(y^{ud}_{\Omega_1})_{jl}}{M^2_{\Omega_{1}}}$	 \\[0.4cm]
$\Phi$ &		&	$-\dfrac{(y^{dq}_{\Phi})_{lk}^*(y^{qu}_{\Phi})_{ij}}{M^2_{\Phi}}$	 \\[0.4cm]
  \hline
 \end{tabular}
% \end{footnotesize}
\caption{Field-by-field contribution to four-fermion operators. }\label{tab:UVmodels3}
%\end{center}
\end{table}
Although all possible couplings of $\mathcal{G}_1$, $\mathcal{H}$,
$\mathcal{Q}_1$ and $\mathcal{Q}_5$ contribute to the operators of
interest (that is why there are no ellipsis), they also induce other
operators; see Ref.~\cite{deBlas:2017xtg} for further details. 

The contributions of each field to the contact-interactions studied in this article after integration out can be found in Tabs.~\ref{tab:UVmodels1}, \ref{tab:UVmodels2} and \ref{tab:UVmodels3}.
Let us also remark that, in the presence of several fields in the Lagrangian, the coefficient of every dimension-six operator is the sum of the contributions of each field.

In the following, we will consider various UV completions, and compare the reach of the limits obtained in this paper with respect to the one of other searches. To start with, let us focus on new scalars. A particularly interesting example is $\Pi_7$, because it does not generate operators others than the ones studied in this article. However, this scenario is already quite bounded by measurements of $\mathcal{B}(B_s\rightarrow s \mu\mu)$. A more interesting example is $\omega_1$ with couplings
\begin{figure}[t]
 \includegraphics[width = 0.3295\columnwidth]{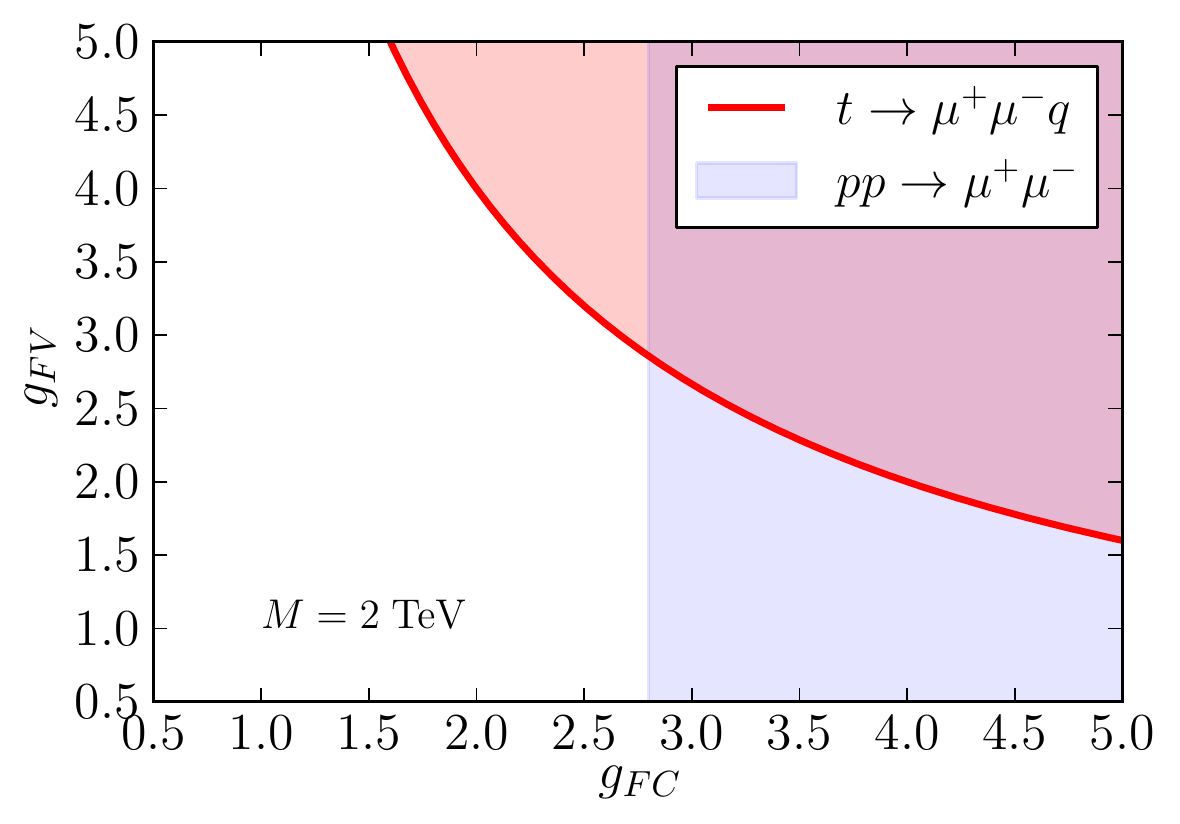}
\includegraphics[width = 0.3295\columnwidth]{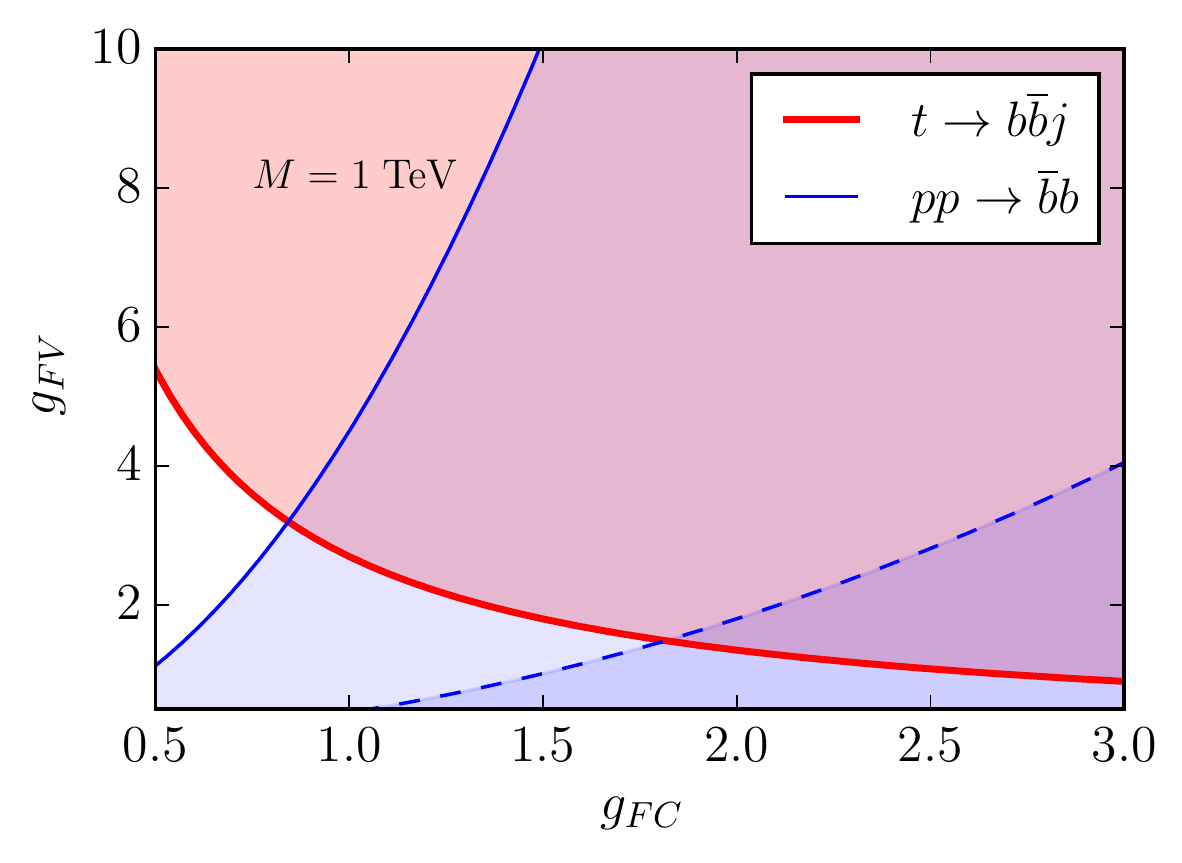}
 \includegraphics[width = 0.3295\columnwidth]{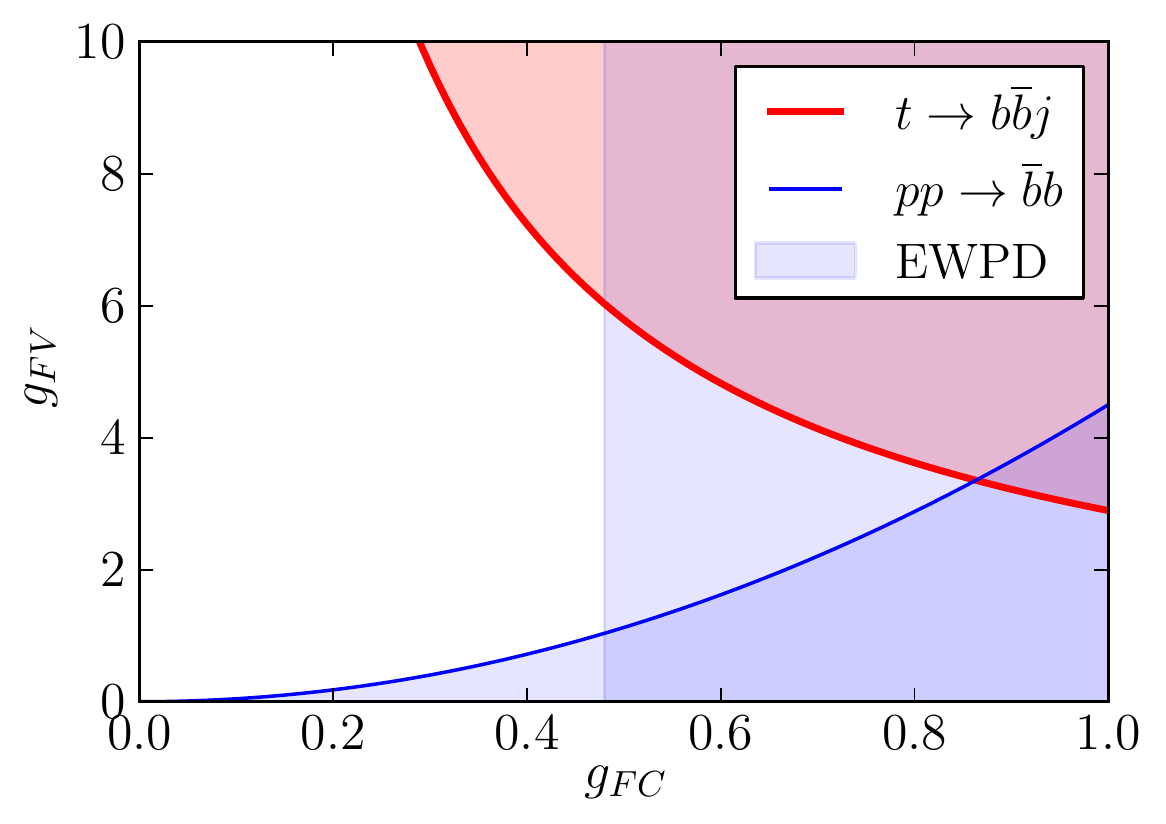}
 \caption{Parameter space region that can be bounded by rare top decays in the 
HL-LHC (red) versus the one that could be excluded using other searches (blue) 
\mc{neglecting systematic uncertainties}. In the left panel 
we consider a scalar leptoquark extension of the SM, in the center and the 
right panels we consider two versions of a $Z^\prime$ model; see text for 
details. }\label{fig:UVbounds}
\end{figure}
\begin{equation}
 L = -\omega_1^\dagger \bigg[ig_{FC}\overline{\mu_R^c} c_R + g_{FV}\overline{\mu_R^c} t_R\bigg] + \text{h.c.}
\end{equation}
At tree level, we obtain $c_{eu}^{2223} = -i g_{FV}g_{FC}/(2 M^2)$, which can be probed in anomalous top decays. On top of it, one also gets $c_{eu}^{2222} = g_{FC}^2/(2M^2)$ and $c_{eu}^{2323} = -g_{FV}^2/(2M^2)$. The first one modifies the tail of the invariant-mass di-muon spectrum. (Resonant searches do not apply because the lepto-quark mediates in t-channel.) A naive rescaling of the bounds in Ref.~\cite{deBlas:2013qqa} with the larger energy, luminosity and smaller PDFs for charms with respect to valence quarks gives an estimated bound at the HL-LHC of $c_{eu}^{2222}\sim 1$ TeV$^{-2}$. The second operator, instead, can be bounded from EWPD. Being the coefficient negative, however, implies that the corresponding bound is very weak~\cite{deBlas:2015aea}. The comparison between the reach of the different searches in this example is depicted in the left panel of Fig.~\ref{fig:UVbounds} for $M=2$ TeV. For smaller masses the pair-production cross section is large enough to better test them in direct production~\cite{Dorsner:2018ynv,CMS-PAS-EXO-17-009}.

Let us now consider the case of the $Z^\prime$, complete singlet of the SM gauge group, with mass $M = 1$ TeV and couplings
\begin{equation}
 L = Z_\mu^\prime \bigg[ g_{FC}\,\overline{b_R}\gamma^\mu b_R + \left\lbrace g_{FV}\,\overline{t_R}\gamma^\mu c_R + \text{h.c.} \right\rbrace \bigg] + \cdots
\end{equation}
After integrating it out, we obtain $c_{ud}^{1(2333)} = g_{FC} g_{FV}$, wich is constrained to be $\lesssim 2.7$ according to Table~\ref{tab:hadbounds}. On the other hand, the $Z^\prime$ can be directly produced in $pp$ collisions initiated by $b$ quarks. The theoretical cross section for this process at the LHC14 is around $\sim 2$ pb for $g_{NC} = 1$. The branching ratio into two $b$-quarks is approximately given by $\mathcal{B}(Z'\rightarrow b\overline{b})\sim g_{FC}^2/(g_{FC}^2 + 2g_{FV}^2)$. Resonant searches at the LHC13 impose a bound on $\sigma(pp\rightarrow Z^\prime)\times\mathcal{B}(Z^\prime\rightarrow b\overline{b})$ of around 0.5 pb~\cite{Aaboud:2018tqo}. A simple rescaling with the energy and luminosity enhancement shows that cross sections ten times smaller could be probed in the HL-LHC. The corresponding bounds on the $g_{FC}$--$g_{FV}$ plane are depicted in the central panel of Fig.~\ref{fig:UVbounds}. (To the best of our knowledge, the current uncertainties in measurements of the single top production cross section make the corresponding bounds not significant.)

The shaded blue region to the right of the thin dashed line assumes $\mathcal{B}(Z^\prime\rightarrow\text{invisible}) = 90~\%$. This case arises for example in models with fermionic dark matter. Ref.~\cite{Chala:2015ama} provides prospects for probing such invisible decays in monojet searches at the HL-LHC. Given their results, which assume that the $Z^\prime$ couples predominantly to light quarks,  it is unlikely that the invisible channel can trigger any sensible bound in our case. Moreover, the corresponding bound from $b\overline{b}$ searches becomes much weaker; see the dashed blue line in the same plot. It is therefore apparent that top FCNCs can provide complementary bounds in  the strongly couple regime.

We repeat the previous exercise for a $Z^\prime$ with couplings
\begin{equation}
  L = Z_\mu^\prime \bigg[ g_{FC}\,\overline{q_L^3}\gamma^\mu q_L^3 + \left\lbrace g_{FV}\,\overline{t_R}\gamma^\mu c_R + \text{h.c.} \right\rbrace \bigg] + \cdots
\end{equation}
Upon integration, we get $c_{qu}^{1(3332)} = g_{FC}g_{FV}$, as well as $c_{qq}^{1(3333)} = g_{FC}^2/2$. This latter operator does not induce FCNCs, but it is constrained by EWPD. Inded, it renormalizes the operators $\mathcal{O}_{\phi q}^{(1)}$ and $\mathcal{O}_{\phi q}^{(3)}$, 
\begin{equation}
 \frac{d(c_{\phi q}^{(1)} + c_{\phi q}^{(3)})}{d\log{\mu}}\sim \frac{2 N_c}{(4\pi)^2}  y_t^2 c_{qq}^{1(3333)} ~,
\end{equation}
which in turn modify the $Z\overline{b_L} b_L$ coupling. Ref.~\cite{deBlas:2015aea} reports a bound of $c_{qq}^{1(3333)}/\Lambda^2 \in [-0.58,0.23] $. Again, we have also constrains from $b\overline{b}$ resonances, as well as from $t\overline{t}$ resonances; the latter being of similar reach. They are all shown in Fig.~\ref{fig:UVbounds} right.

\section{Conclusions}
\label{sec:conclusions}
Using the latest experimental search for $t\rightarrow Zj$, we have obtained the best collider limits on four-fermion operators leading to non-resonant $t\rightarrow \ell^+\ell^- j$, including lepton-flavour-conserving as well as lepton-flavour-violating interactions. We have also shown that, for several operators, our bounds improve over indirect limits from low energy experiments. We have also developed modified versions of current analyses with better reach to the aforementioned dimension-six operators. We have shown that scales of about $\sim 2$ $(3.5)$ TeV can be probed at the HL-LHC for couplings of order $\sim 1$ ($\sim \sqrt{4\pi}$). They are around a factor of $30~\%$ more stringent than the projected bounds using current searches. In light of these results, we urge the experimental collaborations to extend current analyses with signal regions outside the $Z$ peak.

On another front, we have explored the HL-LHC reach to contact-interactions giving $t\rightarrow b\overline{b} j$. We have developed a specific analysis tailored to the kinematic of this process. We have shown that the bounds on such operators using rare-top decays can shed light on the strongly couple regime of the UV. Finally, we have also singled out all possible weakly-couple and renormalizable extensions of the SM that can generate the operators above at tree level. We have selected several of them and shown that large regions of their parameter spaces can be better tested using rare top decays than other observables.\\

\noindent

\textbf{Note added:} During the last stage of this work, 
Ref.~\cite{Gottardo:2018ptv} appeared on the arXiv.
The latter provides bounds on an incomplete set of charged LFV four-fermion 
operators in top decays using a BDT analysis based on $\mathcal{L} = 79.8$ 
fb$^{-1}$. Focusing on the $e\mu$ channel, this reference reports an expected 
bound of $\mathcal{B}(t\rightarrow e\mu q) < 4.8^{+2.1}_{-1.4}\times 10^{-6}$. 
Rescaling to this luminosity, \mc{our expected bounds translate to a bound of 
$\mathcal{B}(t\rightarrow e\mu q) \lesssim 2\times 10^{-5}$.} However, we 
consider also non-LFV decays, as well as a full basis of the SMEFT. We strongly 
encourage the experimental collaborations to adopt the master 
equation~\ref{eq:lepwidth} in this respect.

\section*{ Acknowledgments}%\\[-3mm]
\noindent

We thank Nuno Castro, Gauthier Durieux, Sebastian J\"ager, Alexander Lenz, Adriano Lo Presti, Andreas Meier and Jakub Scholtz for useful discussions. M. C. is funded by the Royal Society under the Newton International Fellowship program. The work of J. S. has been supported by the Spanish MICINN project FPA2013-47836-C3-2-P, the MINECO project FPA2016-78220-C3-1-P (Fondos FEDER) and the Junta de Andaluc\'ia grant FQM101.

\bibliographystyle{style}
\bibliography{notes}{}

\end{document}